\documentclass[aps, superscriptaddress, floatfix, 12pt]{revtex4}
\usepackage{graphicx}
\usepackage{amsmath}
\usepackage{amssymb}

\newcommand{\be}{\begin{eqnarray}}
\newcommand{\ee}{\end{eqnarray}}
\newcommand\del{\partial}
\newcommand{\el}{\nonumber \hfill \\}
\newcommand{\Tr}{\mathrm{Tr}}
\newcommand{\nn}{\nonumber }

\begin{document}

\title{Volume Dependence of the Pion Mass in the
Quark-Meson-Model}
\author{J. Braun}
\affiliation{Institute for Theoretical Physics, University of
  Heidelberg, Philosophenweg 19, 69120 Heidelberg}
\author {B. Klein} 
\affiliation{Institute for Theoretical Physics, University of
  Heidelberg, Philosophenweg 19, 69120 Heidelberg}
\author{H.J. Pirner}
\affiliation{Institute for Theoretical Physics, University of
  Heidelberg, Philosophenweg 19, 69120 Heidelberg}
\affiliation{Max-Planck-Institut f\"ur Kernphysik, Saupfercheckweg 1,
  69117 Heidelberg}

\date{\today}

\begin{abstract}
We consider the quark-meson-model in a finite three-dimensional volume
using the Schwinger proper-time
renormalization group. We derive and solve the flow equations 
for finite volume  in local potential
approximation. In order to break chiral
symmetry in the finite volume, we introduce a small current quark
mass. The corresponding 
effective meson potential breaks chiral $O(4)$ symmetry explicitly,
depending on $\sigma$ and $\vec{\pi}$ fields separately. 
We calculate the volume dependence of the pion mass and of the
pion decay constant with the renormalization group flow equations and
compare with recent 
results from chiral 
perturbation theory in a finite volume.
\end{abstract}

\maketitle

\section{Introduction}

The study of QCD in a finite volume has been of interest for
quite some time.
Accurate results of
lattice simulations with dynamical fermions 
necessitate
understanding  finite volume effects. A variety of different
methods has been proposed, cf. refs. \cite{Becirevic:2003wk,
  AliKhan:2003cu, Guagnelli:2004ww, Arndt:2004bg, Procura:2003ig,
  Beane:2004rf, Berbenni-Bitsch:1997fj,
  Colangelo:2004xr, Colangelo:2003hf},  to  
extrapolate reliably from finite lattice volumes to the infinite volume.
Finite volume partition functions for QCD have attracted interest in
their own right, because they allow an
exact description of QCD at low energies 
\cite{Gasser:1986vb, Gasser:1987ah, Gasser:1987zq, Leutwyler:1992yt}.
The low energy behavior of QCD is determined by spontaneous chiral symmetry
breaking \cite{Gasser:1983yg},
which, however, does not occur in a finite volume.
If the current quark mass is set equal to zero, in a finite volume the
expectation value for the order parameter of chiral symmetry breaking
vanishes, remaining zero even for arbitrary large volumes. The order
parameter has a finite expectation value only when the
infinite volume limit is taken before the quark mass is set
to zero. 

The box size $L$, the pion mass $m_\pi$ and the pion decay constant
$f_\pi$ are the relevant scales  
for the transition between the regimes with a
strongly broken and with an
effectively restored chiral symmetry \cite{Gasser:1986vb}. As a measure
of explicit symmetry breaking, the pion mass is of particular
importance. 
It is primarily the dimensionless product 
$m_\pi L$ that determines in which
regime the system exists for a given pion mass and volume. 
In order to study chiral symmetry breaking in a
finite volume, it is essential to introduce a finite quark
mass as a parameter that explicitly breaks the chiral symmetry.
Such an explicit symmetry breaking is quite natural in theories
which involve effective chiral Lagrangians. 

QCD at low energy can be studied by a wide variety of approaches which
in essence all rely on the same fact: Due to  spontaneous breaking of
chiral symmetry, low-energy QCD is dominated by  massless
Goldstone bosons associated with the broken symmetry. Since these
Goldstone bosons interact only weakly, the low-energy limit of QCD can be
described in terms of an effective theory of these fields. 
A description in terms of effective chiral Lagrangians becomes
even better if one considers the partition 
functions in finite Euclidean volume. Compared to the light degrees of
freedom, contributions of heavier particles are suppressed by $e^{-ML}$, 
where $M$ is the typical separation of the hadronic mass 
scale from the Goldstone masses. This separation of mass scales is at the
origin of the description of QCD with effective theories in terms of the
light degrees of freedom only. 

Groundbreaking work has been done by Gasser and Leutwyler
\cite{Gasser:1986vb, Gasser:1987ah, Gasser:1987zq} 
in chiral perturbation theory, and by Leutwyler and 
Smilga for the eigenvalue spectrum of the QCD Dirac operator
\cite{Leutwyler:1992yt}. Random 
matrix theory \cite{Shuryak:1992pi,
  Verbaarschot:1994qf} predicts analytically the 
volume and quark mass dependence of the chiral condensate
\cite{Verbaarschot:1995yi, Damgaard:1999dr, Damgaard:1999tk}, and
the eigenvalue 
spectrum which has been well 
confirmed by numerous lattice results, see
eg. \cite{Berbenni-Bitsch:1997tx, Damgaard:1998ie, Damgaard:1999tk}.  
Such analytic predictions have been extremely useful as a check for
calculations in lattice gauge theory.  

In the context of the
renormalization group (RG) approach, most calculations so far have been
done in a chirally symmetric formulation \cite{Schaefer:1999em,
  Meyer:2001zp, Papp:1999he, Schaefer:2004en}. In such a framework,
explicit symmetry breaking terms do not affect the renormalization
group flow. This means that 
pion contributions to the renormalization group flow come from exactly
massless pions. Small quark masses can be converted into a 
small linear term in the meson potential which is not renormalized and
can therefore be added at the end of the evolution.
In this paper, we will present a formulation which includes explicit symmetry
breaking in the renormalization group flow and which can be used also for
a finite volume.

We use the chiral quark-meson-model, which contains a scalar sigma
meson and quarks in addition to the pion degrees of freedom.
It is well known that the linear sigma model alone is not compatible
with the low energy $\pi \pi$-scattering data \cite{Gasser:1983yg}.
Due to the presence of quarks, the
low energy constants of chiral perturbation theory are reproduced
\cite {Jendges, Jungnickel:1997yu}. 
The quark-meson-model is evolved with 
renormalization group flow equations which connect different scales. 
As we will
demonstrate, the inverse box size $1/L$ acts as an effective cutoff
scale which freezes the evolution when the evolution parameter
$k<\pi/L$.  In the same way as the renormalization group flow equations
describe the dependence of the results on the renormalization cutoff
scale $k$, they also describe the dependence on the additional scale
imposed by the finite volume. The summation of higher loop graphs in
this approach does allow to extend the calculation to smaller quark
masses and volumes, where perturbative calculations
lack convergence. Deviations of the 
calculated pion mass correction in the finite volume from chiral 
perturbation theory are
found in this region.  Our results agree, however, in the case of
large pion masses and large volumes. 

The paper is organized as follows:
In section~\ref{sec:chPT}, we review  recent results from the application
of chiral perturbation theory to the problem of finite volume effects.
In section~\ref{sec:RGflow}, we show how the evolution equation for the 
effective meson potential is modified in a  finite
volume. Details of the numerical evaluation are discussed in
section~\ref{sec:numerical}. In section~\ref{sec:results} we
present our final results, which are discussed  
in section~\ref{sec:conclusions}.

\section{Finite Volume effects in chiral perturbation theory}
\label{sec:chPT}
For finite volume, 
massless Goldstone bosons dominate the action of a theory with broken 
chiral symmetry. 
In chiral perturbation theory (chPT)
\cite{Gasser:1983yg},  
the pion mass, the pion decay constant and the chiral
condensate have been calculated 
\cite{Gasser:1987ah, Gasser:1987zq, Gasser:1986vb}. 
The expansion parameters are the
magnitude of the three-momentum $|\vec{p}|$ and the mass of the pion
$m_\pi$ as the lightest degree of freedom compared with
the chiral symmetry breaking scale $4 \pi f_\pi$.

Depending on the size $L$ of the volume and the pion mass $m_\pi$,
chiral perturbation theory distinguishes  between two
different power counting schemes. If the size of
the box is much larger than the Compton wavelength of the pion 
$L \gg 1/m_{\pi}$, the lowest nonzero pion momentum 
is smaller than the pion mass ($p_{\mathrm{min}} \sim \frac{2 \pi}{L} \ll
m_\pi$) and the normal power counting
scheme applies (``$p$-regime''). In
this case, the pions 
are constrained very little by the presence of the box, and finite
size effects are comparatively small \cite{Gasser:1987zq}. 
If, on the other hand,
the size of the box is 
smaller than the Compton wavelength of the pion, the normal
chiral expansion breaks down, since the smallest momentum 
$p_{\mathrm{min}} \sim
\frac{2\pi}{L} \gg m_\pi$ is now much larger than the pion mass
(``$\epsilon$-regime''). 
In this case, the partition function is dominated by the zero modes.
After solving the zero-momentum sector of the theory exactly,
one expands the finite momentum modes to one-loop order.

A very useful tool to study the effects of a finite volume on
the mass of the pion is L\"uscher's formula \cite{Luscher:1985dn}.
It relates the leading finite volume corrections for the pion mass
in Euclidean volume to the $\pi \pi$-scattering amplitude in infinite
volume. Corrections to the leading order behavior 
drop at the least as ${\mathcal O}(e^{- \bar m L})$ where $\bar m \ge
\sqrt{3/2} m_\pi$. 
For the particular case of the pion mass, the formula for the relative 
deviation $R[m_{\pi}(L)]$ of the pion mass $m_\pi(L)$ 
in the finite volume from the
pion mass in the infinite volume $m_\pi(\infty$ ) reads as follows:
\be
R[m_{\pi}(L)] &=& \frac{m_\pi(L) - m_\pi(\infty)}{m_\pi(\infty)}\el 
               &=& -\frac{3}{16 \pi^2}\frac{1}{m_\pi} \frac{1}{m_\pi L} 
                   \int_{-\infty}^{\infty} dy\, 
F(i y) e^{-\sqrt{m_\pi^2+y^2} L}+{\mathcal O}(e^{-\bar m L}).
\label{eq:Luscher}
\ee
$F$ is the forward $\pi \pi$-scattering
amplitude as a function of the energy variable $s$ continued to 
complex values.
New results \cite{Colangelo:2003hf, Colangelo:2004xr}
have recently been obtained by combining  L\"uscher's formula with 
a calculation of the scattering amplitude in chiral perturbation theory.
A next-to-next-to leading order calculation of $F$ alone 
does not seem to give a reliable and satisfactory result.
A one-loop calculation using L\"uscher's formula gives a shift in the pion
mass, for example, which is substantially lower than the one expected
form the full one-loop calculation in chiral perturbation theory as
performed by Gasser and Leutwyler \cite{Gasser:1987ah}. 
This estimate of the finite volume
effects can be improved, if one uses the mass correction obtained from
L\"uscher's formula with a $\pi \pi$-scattering
amplitude including higher orders to correct 
the full one-loop chiral perturbation theory result
\cite{Colangelo:2003hf}. 
Since L\"uscher's formula has sub-leading corrections
${\mathcal O}\left(\exp[-\sqrt{3/2} m_\pi L]\right)$, the corrections to
the leading result increase for decreasing pion mass at fixed volume
size $L$ with $m_\pi L$.  
As pointed out by the authors in
\cite{Colangelo:2003hf}, the L\"uscher formula becomes a less reliable
approximation exactly for those values of the pion mass for which
the chiral expansion converges especially well.

The renormalization group flow equations do not rely on 
either box size or pion
mass as an expansion
parameter, and do not require to distinguish between two different
regimes. They remain valid as long as the lowest momenta and the
masses of the heaviest particles remain
below the ultraviolet cutoff scale $\Lambda_{UV} \approx 1.5$ GeV.
The beauty of the renormalization group method is precisely that 
the flow equations connect different scales. 
In the same way as the renormalization group flow equations 
describe the dependence of the results on the infrared cutoff
scale $k$, they also describe the dependence on the additional scale
imposed by the finite volume.

\section{Renormalization group flow equations for the Quark-Meson-model}
\label{sec:RGflow}

The quark-meson model is an $SU(2)_{L}\otimes SU(2)_{R}$ invariant
linear $\sigma$-model with chiral mesons $\phi=(\sigma,\vec{\pi})$
coupled to constituent quarks $q$.
It is an effective model for dynamical spontaneous chiral 
symmetry breaking
at intermediate scales of $k \lesssim \Lambda_{UV}$, where
the ultraviolet scale $\Lambda_{UV} \approx 1.5\;\mathrm{GeV}$ 
is determined by the validity of a hadronic representation of QCD. 
At the UV scale $\Lambda_{UV}$, the quark-meson-model is defined by
the effective action
\be
\Gamma_{\Lambda_{UV}}[\phi]&=& \int d^4 x
\left\{ \bar{q} \gamma{\partial}q + g\bar{q}(\sigma+i\vec{\tau}\cdot\vec{\pi}
\gamma_{5}) q +
\tfrac{1}{2}(\partial_{\mu}\phi)^{2}+U(\phi)  
+m_c\bar{q}q\right\} 
\ee
in a four-dimensional Euclidean volume with compact Euclidean time direction.
The partition function $Z$ has the path integral representation 
\begin{align}
Z 
& =\int\! {\mathcal D}\bar{q}\!\int\! {\mathcal D}q\!\int\!
 {\mathcal D}\phi \exp\left(- \Gamma_{\Lambda_{UV}} \right),
\end{align}
 where in the Euclidean time direction periodic and anti-periodic
 boundary conditions apply for bosons and fermions, respectively. 
A Gaussian approximation to the path integral
followed by a Legendre transformation yields the one-loop effective
action for the scalar fields $\phi$, 
\be
\Gamma[\phi]&=& \Gamma_{\Lambda_{UV}}[\phi] -
\Tr\log\left(\Gamma_{F}^{(2)}[\phi]\right) +
\frac{1}{2}\Tr\log\left(\Gamma_{B}^{(2)}[\phi]\right) 
\ee
where $\Gamma_{B}^{(2)}[\phi]$ and $\Gamma_{F}^{(2)}[\phi]$ are the
inverse two-point functions for the bosonic and fermionic fields,
evaluated at the vacuum expectation value of the mesonic
field $\phi$.  Here, the boundary conditions of the functional
integral appear in 
the momentum traces and we neglect contributions from mixed
 quark-meson-loops. 
In order to regularize the functional traces, we use the Schwinger
proper time representation of the logarithms.
We consider the effective action $\Gamma$ in a local
potential approximation (LPA), which represents the lowest order in
the derivative expansion and incorporates fermionic as well as bosonic
contributions to the potential density $U$. In this approximation, the
effective field $\phi$ is considered to be constant over the entire 
volume.

The scale dependence is
introduced through the infrared cutoff function $f_{a}(\tau k^{2})$, which
regularizes the Schwinger proper time integral. A cutoff function of
the form
\be
k\frac{\partial}{\partial k} f_{a}(\tau k^{2}) & = &
-\frac{2}{\Gamma(a+1)}(\tau k^{2})^{a+1}e^{-\tau
k^{2}}\label{eq:cutoff-fct}
\ee
satisfies the required regularization conditions
\cite{Schaefer:1999em, Meyer:1999bz, Papp:1999he, Meyer:2001zp}. 
We obtain a renormalization group flow equation for the effective
potential by performing  a renormalization group improvement and
replacing the bare two-point functions by the renormalized,
scale-dependent two-point functions, 
\be
k \frac{\partial}{\partial k} \Gamma_k[\phi] &=& \frac{1}{2}
\Tr \int_{0}^{\infty} \frac{ d \tau}{\tau} \left[ k
\frac{\partial}{\partial k} f_a(\tau k^2)\right] \exp[ -\tau
 \Gamma_{B,k}^{(2)}[\phi]]\el
& & - \Tr \int_{0}^{\infty} \frac{ d \tau}{\tau} \left[ k
\frac{\partial}{\partial k} f_a(\tau k^2)\right] \exp[ -\tau
  \Gamma_{F,k}^{(2)}[\phi]] 
\ee
In LPA, the effective action reduces to the effective potential
through the relation
\be
\Gamma_k[\phi] &=& \int d^4 x \; U_k(\phi).
\ee
The renormalization group improved evolution equation 
for the effective potential in infinite volume is given 
by 
\be
\label{eq:CutoffAbl}
k \frac{\partial}{\partial k}U_k(\phi,L \to \infty, T) &=& \frac{1}{2}
\int_{0}^{\infty} \frac{ d \tau}{\tau} \int \frac{d^4 p}{(2
  \pi)^4} \Big\{ 4 N_c N_f \exp[-\tau(p^2+ M_q^2(\sigma,\vec{\pi}^{2}))]  \\
& & - \exp[-\tau (p^2 + M_{\sigma}^2(\phi^{2}))]
    - 3 \exp[-\tau (p^2 + M_{\pi}^2(\phi^{2}))] \Big\} 
k\frac{\partial}{\partial k} f_a(\tau k^2).\nn 
\end{eqnarray}
It is necessary to choose the parameter $a$ in such a way that the
resulting integrals over the proper time parameter $\tau$ remain
finite. In infinite volume, the lowest possible integer value is $a=2$. 
The diagonalization of the meson mass matrix gives the running
meson masses which depend on the effective potential. Without explicit
symmetry breaking, they are of the form
\be
M_{\sigma}^2 &=&2 \frac{\partial U_k}{\partial \phi^2}+ 4  \phi^2 
\frac{\partial^2
U_k}{(\partial \phi^2)^2}\\
M_{\pi}^2 &=& 2 \frac{\partial U_k}{\partial \phi^2}.
\ee
To derive renormalization group flow equations 
in the finite volume $L^{d-1}$ at finite temperature $T$,
we replace the integrals over the momenta by a sum
\begin{equation}
\int
dp_{i}\,\ldots\rightarrow\frac{2\pi}{L}\sum_{n_i=-\infty}^{\infty}\ldots\,
\end{equation}
and 
apply periodic boundary conditions for bosons and anti-periodic boundary
conditions for fermions in time and space-like directions. The sums
run from $-\infty$ to $+\infty$, where the vector $\vec{n}$ denotes
$(n_{1},n_{2},...,n_{d-1})$. The Matsubara frequencies take the value
$\omega_{n}=2\pi nT$ for bosons and $\nu_{n}=(2n+1)\pi T$ for fermions,
respectively. In the following we use the short-hand notation
\be
p_{F}^{2}&=&\sum_{i=1}^{d-1}p_{i}^{2} = \frac{4\pi^{2}}{L^{2}}
\sum_{i=1}^{d-1}\left( n_{i}+\frac{1}{2}\right)^{2} \\
p_{B}^{2}&=&\sum_{i=1}^{d-1}p_{i}^{2} = \frac{4\pi^{2}}{L^{2}}
\sum_{i=1}^{d-1} n_{i}^{2};\quad
\ee
for the momenta of the fermions and bosons, respectively.
In finite volume,
we allow explicit symmetry breaking in the effective potential, which then
becomes a function of two variables $\sigma$ and $\vec{\pi}^{2}$
separately. 
The corresponding expression to eq.~(\ref{eq:CutoffAbl}) is
\be
k\frac{\partial}{\partial k}U_{k}(\sigma,\vec{\pi}^{2},L,T)\! & = &
\frac{1}{2}\frac{T}{L^{d\!-\!1}}\int\!\frac{d\tau}{\tau}
\!\sum_{l}\!\sum_{\vec{n}}\Big\{4 N_{c}  N_{f} \exp\left[{-\!\tau(
    \nu_{l}^{2}+p_F^{2}+M_{q}^{2}(\sigma,\vec{\pi}^{2})\!)}\right]\nn \\
 &  & \quad -\sum_{i=1}^{4}\exp\left[-\!\tau(
 \omega_{l}^{2}+p_B^{2}+M_{i}^{2}(\sigma,\vec{\pi}^{2})\!)\right]\Big\}
 k\frac{\partial}{\partial k} f_{a}(\tau k^{2}).
\ee
The $M_{i}^2$, $i= 1, \ldots, N_f^2$, $N_f=2$, are the eigenvalues of
the second derivative matrix 
\be
\left(U_k (\sigma, \vec{\pi}^2)\right)^{ij} = \frac{\del^2 U_k}{\del
  \phi_i \del \phi_j} 
\ee
of the meson potential $U_k(\sigma, \vec{\pi}^2)$ with respect to the
fields 
$\phi=(\sigma, \vec{\pi})$. 
They depend only on the magnitude of the pion fields
$\vec{\pi}^2$ and are independent of the direction. 
We wish to stress the importance of this point, 
since otherwise the meson contributions from the flow equations 
are not compatible with the ansatz for the potential which we will
introduce below.

The second derivative matrix is given by
\be
\left(
\begin{array}{cccc} 
U_{\sigma\sigma} & U_{\vec{\pi}^2 \sigma} 2 \pi^{(1)} & U_{\vec{\pi}^2
  \sigma} 2\, \pi^{(2)} & U_{\vec{\pi}^2 \sigma} 2 \,\pi^{(3)} \\
U_{\sigma\vec{\pi}^2} 2 \pi^{(1)} & 2\, U_{\vec{\pi}^2} + U_{\vec{\pi}^2
  \vec{\pi}^2} 4\, (\pi^{(1)})^2 &  U_{\vec{\pi}^2
  \vec{\pi}^2} 4\, \pi^{(1)} \pi^{(2)} & U_{\vec{\pi}^2
  \vec{\pi}^2} 4\, \pi^{(1)} \pi^{(3)}  \\ 
U_{\sigma\vec{\pi}^2} 2\, \pi^{(2)} &  
U_{\vec{\pi}^2  \vec{\pi}^2} 4\, \pi^{(2)} \pi^{(1)} & 
2\, U_{\vec{\pi}^2} + U_{\vec{\pi}^2 \vec{\pi}^2} 4\, (\pi^{(2)})^2 & 
U_{\vec{\pi}^2  \vec{\pi}^2} 4 \,\pi^{(2)} \pi^{(3)} \\
U_{\sigma\vec{\pi}^2} 2\, \pi^{(3)} &
U_{\vec{\pi}^2  \vec{\pi}^2} 4\, \pi^{(3)} \pi^{(1)} &
U_{\vec{\pi}^2  \vec{\pi}^2} 4\, \pi^{(3)} \pi^{(2)} &
2 U_{\vec{\pi}^2} + U_{\vec{\pi}^2 \vec{\pi}^2} 4\, (\pi^{(3)})^2 
 \\
\end{array}
\right)
\ee
where we have suppressed the scale index $k$ of the potential and use the
abbreviations 
\be
U_{\sigma} = \frac{\partial U}{\partial \sigma},\;\; U_{\pi^{(a)}} =
\frac{\partial U}{\partial \vec{\pi}^2} \frac{\partial
  \vec{\pi}^2}{\partial \pi^{(a)}}=U_{\vec{\pi}^2} 2 \pi^{(a)},
\;\;U_{\vec{\pi}^2} = 
\frac{\partial U}{\partial \vec{\pi}^2},
\ee
and the corresponding expressions for the higher derivatives.
The eigenvalues of this matrix are given by
\be
M_1^2&=& \frac{1}{2} \left[ 2\, U_{\vec{\pi}^2} + 4 \, \vec{\pi}^2 \,
  U_{\vec{\pi}^2 \vec{\pi}^2} + U_{\sigma \sigma} +
  \sqrt{(2\, U_{\vec{\pi}^2} + 4\, \vec{\pi}^2
      U_{\vec{\pi}^2 \vec{\pi}^2} - U_{\sigma \sigma} )^2 
          + 16\, \vec{\pi}^2\, U_{\sigma \vec{\pi}^2}^2
   } \right], 
\el
M_2^2&=& 2 \, U_{\vec{\pi}^2},\;\;\;M_3^2 = 2 \, U_{\vec{\pi}^2}, \el
M_4^2&=&  \frac{1}{2} \left[ 2\, U_{\vec{\pi}^2} + 4 \, \vec{\pi}^2 \,
  U_{\vec{\pi}^2 \vec{\pi}^2} + U_{\sigma \sigma} -
  \sqrt{(2\, U_{\vec{\pi}^2} + 4\, \vec{\pi}^2
      U_{\vec{\pi}^2 \vec{\pi}^2} - U_{\sigma \sigma} )^2 
          + 16\, \vec{\pi}^2\, U_{\sigma \vec{\pi}^2}^2
   } \right]. 
\ee
For vanishing cross terms $U_{\sigma \vec{\pi}^2}$, the last eigenvalue
reduces to $2\, U_{\vec{\pi}^2} + 4 \, \vec{\pi}^2 \, U_{\vec{\pi}^2
  \vec{\pi}^2}$, which corresponds to a derivative in ``radial''
direction in the pion-subspace. 
Especially for $\vec{\pi}^2=0$, the three pion modes have equal
masses.
We also note that the pion
fields appear only in the combination $\vec{\pi}^2$ in the
eigenvalues, despite the fact that the derivative matrix contains
terms linear in $\pi^{(a)}$. The reason is that the rotational symmetry of
the pion space remains unbroken even in the presence of  explicit
symmetry breaking terms in the sigma direction. 

As discussed in \cite{Braun:2003ii}, we are able to perform the sum over
the thermal Matsubara frequencies analytically and obtain an evolution
equation which contains the
Fermi-Dirac-distribution $n_{F}(E)$
and the Bose-Einstein-distribution $n_{B}(E)$, 
\begin{eqnarray}
k\frac{\partial}{\partial k} U_{k}(\sigma,\vec{\pi}^{2},L,T)\! 
& = & 
\!\frac{(-1)^{a}}{2 \Gamma(a+1)}\frac{k^{2(a+1)}}{L^{d\!-\!1}} 
\frac{\partial^{a}}{(\partial k^{2})^{a}}\sum_{\vec{n}}
\Big(\sum_{i=1}^{4}\frac{1}{E_{i}}(1+2 n_{B}(E_{i}))\nonumber \\
&  &
\qquad\qquad-\frac{4 N_{c} N_{f}}{E_{q}}(1-2 n_{F}(E_{q}))\Big)\,, 
\label{eq:FGVOLALLGT}
\end{eqnarray}
with
\begin{equation}
n_{F}(E) = \frac{1}{e^{\frac{E}{T}}+1}\,, \; \; n_{B}(E) =
\frac{1}{e^{\frac{E}{T}}-1}\,, 
\end{equation}
where in the absence of a chemical potential for the fermions the
Fermi-Dirac-distributions for quarks and antiquarks coincide. 
The effective energies are defined by
\begin{equation}
E_{q}^{2}=k^{2}+p_{F}^{2}+M_{q}^{2}(\sigma,\vec{\pi}^{2});\quad
E_{i}^{2}=k^{2}+p_{B}^{2}+M_{i}^{2}(\sigma,\vec{\pi}^{2}).
\end{equation}
The parameter $a$ in the cutoff function is given by $a=2$ in the
finite volume, in order to ensure that we can compare with the results
in infinite volume.
In the limit of vanishing temperature $T=0$, we find in $d=4$ the finite
volume evolution equation of the meson potential:
\be
k\frac{\partial}{\partial
  k} U_{k}(\sigma,\vec{\pi}^2,
L)&=&\frac{3}{16}\frac{k^{6}}{L^{3}}\sum_{\vec{n}}\Big(-\frac{4
  N_{c} N_{f}}{(E_{q}(\vec{n},L))^{5}} +\sum_{i=1}^{4}
\frac{1}{(E_{i}(\vec{n},L))^{5}} \Big)\,\label{eq:FGVolA2}
\ee

The polynomial ansatz for the meson potential is determined by the following
idea: Since the current quark mass is the only source of symmetry
breaking, the quark term in the flow equation
determines the symmetry breaking terms of the potential. The
constituent quark mass can be expanded 
around a finite expectation value
of the mesonic fields, which is chosen in the direction of the field
$\sigma$,
\be
M_q^2&=&g^2 [(\sigma + m_c)^2 +\vec{\pi}^2]= g^2 [(\sigma
+\sigma_0-\sigma_0 + m_c)^2
+\vec{\pi}^2] \el
     &=&g^2 [(\sigma_0 + m_c)^2 + 2 m_c (\sigma - \sigma_0) +
  (\sigma^2 +\vec{\pi}^2 -\sigma_0^2)].
\label{eq:mqexpansion}
\ee
We have rescaled $m_c$ by a factor $g$ for convenience, so that the
physical current quark mass is given by $g m_c$.
From this expression, we read off that the contributions to the
potential from the fermionic terms in the flow equations can all be
expressed in terms of powers of the combinations $(\sigma^2
+\vec{\pi}^2 -\sigma_0^2)$ for the symmetric part and $(\sigma -
\sigma_0)$ for the symmetry breaking parts.
Therefore, we make for the meson potential the ansatz
\be
U_k(\sigma, \vec{\pi}^2)&=& \sum_{i=0}^{N_\sigma}
\sum_{j=0}^{[\frac{1}{2}(N_\sigma-i)]} a_{ij}(k) 
(\sigma-\sigma_0)^i(\sigma^2 +\vec{\pi}^2 - \sigma_0^2)^j
\label{eq:potansatz}
\ee
The flow equations for the coefficients in this potential are derived
in the appendix.

Incorporating the explicit breaking of the chiral symmetry into the
potential and the flow from the start has several advantages. 
The  polynomial expansion above evolves
automatically from a potential with small symmetry breaking peaked 
around $\langle \sigma \rangle \approx 0$ to a potential with large 
symmetry breaking peaked at a value $\langle \sigma \rangle \approx f_{\pi}$. 
Without explicit symmetry breaking, the 
polynomial expansion in $\phi^2$ has to be  changed from  
a parametrization in terms of powers of $\phi^2$ to
$(\phi^2-\phi_0^2)$ at the chiral symmetry breaking scale
\cite{Schaefer:1999em}. 

As is well known, a linear symmetry breaking term remains unchanged
in the renormalization group flow \cite{Zinn-Justin:1989mi}. Therefore the
usual strategy is to evolve the
potential without a symmetry breaking term. Explicit symmetry breaking
is then taken into account after the quantum fluctuations have been
integrated out on all scales \cite{Berges:1997eu,
  Jungnickel:1995fp, Schaefer:1999em}. In an
infinite volume and for small quark 
masses, this is perfectly 
acceptable and will yield the correct results.
In a finite volume, however, the situation is different. 
Since chiral symmetry is not spontaneously broken, 
explicit symmetry
breaking has to be included on all scales in the renormalization group
flow to obtain a nonzero value for the order parameter.
Otherwise, divergences from massless Goldstone bosons would restore
the symmetry. 
In this context, we would like to point out that even in the
absence of a symmetry breaking term, the pion decay constant
does not remain zero on all renormalization scales $k$. On some
intermediate scale below the chiral symmetry
breaking scale, $k < k_{\chi}$, where the quantum fluctuations are
only partially 
integrated out, it acquires a nonzero expectation value, and chiral
symmetry is spontaneously broken. However, the emergence of exactly
massless Goldstone bosons dominates the infrared evolution of the 
potential and counteracts the formation of a symmetry breaking
condensate. 

When the potential is expanded in a polynomial in a theory with
exactly massless Goldstone bosons, divergences appear in the flow
equations for the coefficients of operators of mass dimension higher
than four \cite{Papp:1999he}. As an added benefit of including
explicit symmetry 
breaking, the presence of a finite pion mass regulates these
IR divergences. 

\section{Numerical evaluation}
\label{sec:numerical}

We have solved the RG flow equations numerically and
present the results for the volume dependence of
the pion mass and the pion decay constant in the following section. 
For the numerical evaluation, we have used the polynomial ansatz for
the effective potential given in eq.~(\ref{eq:potansatz}), and expanded up
to fourth order in the fields:
\be
U_k(\sigma, \vec{\pi}^2)&=& 
a_{00}(k) 
+ a_{01}(k)(\sigma^2 +\vec{\pi}^2 - \sigma_0^2) 
+ a_{02}(k)(\sigma^2 +\vec{\pi}^2 - \sigma_0^2)^2 \el
&&+ a_{10}(k) (\sigma-\sigma_0)
+ a_{20}(k) (\sigma-\sigma_0)^2 
+ a_{30}(k) (\sigma-\sigma_0)^3
+ a_{40}(k) (\sigma-\sigma_0)^4 \el
&&+ a_{11}(k) (\sigma-\sigma_0)(\sigma^2 +\vec{\pi}^2 - \sigma_0^2) 
\label{eq:numpotansatz}
\ee 
Here, we first discuss our choice of model parameters at the UV scale, and some
details of the numerical evaluation.

%\subsection{Determination of parameters}
The UV scale itself is determined from physical considerations as the
scale below which a description of QCD with hadronic degrees of
freedom is appropriate. Here, we choose $\Lambda_{UV} = 1.5\; \mathrm{GeV}$.
At the ultraviolet scale $\Lambda_{UV}$, the free parameters of the
quark-meson-model are  the meson 
mass $m_{UV}$, the four-meson-coupling $\lambda_{UV}$, and the current
quark mass $g m_c$, which controls the degree of explicit symmetry breaking. 
The Yukawa coupling $g$ does not evolve in the present approximation
\cite{Jungnickel:1995fp, Berges:1997eu, Schaefer:1999em}. 
We choose $g=3.26$, which leads to a reasonable
constituent quark mass of $M_q = g (f_\pi +m_c) \approx 310
\;\mathrm{MeV}$ for physical values for the pion decay constant
$f_\pi=93 \;\mathrm{MeV}$ and the current quark mass $g m_c=7\;\mathrm{MeV}$. 

%%%%%%%%%%%%%%%%%%%%%%%%%%%
\begin{table}
\begin{ruledtabular}
\begin{tabular}{rrrrrr}
%\hline
%\hline
$\Lambda _{UV}\;\; \mathrm{[MeV]}$ & $m_{UV}\;\; \mathrm{[MeV]}$ &
  $\lambda_{UV}$ & $g m_c \; \mathrm{[MeV]}$ &  $f_\pi \;\;
\mathrm{[MeV]}$ & $m_\pi \; 
  \mathrm{[MeV]}$ \\
\hline
1500 & 779.0 & 60 &  2.10 & 90.38 & 100.8\\
1500 & 747.7 & 60 &  9.85 & 96.91 & 200.1\\
1500 & 698.0 & 60 & 25.70 & 105.30 & 300.2\\
%\hline
%\hline
\end{tabular}
\end{ruledtabular}
\caption{\label{tab:start} Values for the parameters 
  at the $UV$-scale used in the numerical evaluation. The parameters
  are determined in infinite volume by fitting to a particular pion
  mass and the 
  corresponding value of the pion decay constant, taken from chiral
  perturbation theory. Note that in our notation, the physical current
  quark mass corresponds to $g m_c$.}
\end{table}
%%%%%%%%%%%%%%%%%%%%%%%%%%

In table~\ref{tab:start}, we summarize the three parameter sets
which we used in obtaining our results for pion masses of
$100$, $200$ and $300\;\mathrm{MeV}$.  
We determine these UV parameters by fitting to a particular value for
the pion mass $m_\pi(\infty)$ and to the corresponding value for the
pion decay constant $f_\pi(\infty)$ in
infinite volume. We then evolve the RG equations with these parameters 
to predict the volume dependence of $f_\pi(L)$ and $m_\pi(L)$. 

For any value of the pion mass, the
corresponding value of the pion decay constant is taken from chiral
perturbation theory \cite{Colangelo:2003hf}. The pion mass is
mainly controlled by the value of the 
current quark mass, which parametrizes the symmetry breaking. The
current quark mass varies from
approximately $2$ MeV for a pion mass of $100\; \mathrm{MeV}$ to
about $10\; \mathrm{MeV}$ for $m_\pi=200\; \mathrm{MeV}$, it has to be
increased to approximately $25\; \mathrm{MeV}$ for $m_\pi=300\;
\mathrm{MeV}$. To achieve the correct corresponding values for the
pion decay constant, the meson mass at the UV scale has to be
decreased from approximately $m_{UV}=780\; \mathrm{MeV}$ to
$m_{UV}=700\; \mathrm{MeV}$, while the  pion mass increases from $100$
to $300\; \mathrm{MeV}$. The four-meson-coupling $\lambda_{UV}$ is fixed.
We have checked that our results are to a very large degree
independent of the particular 
choice of UV parameters: Different sets of parameters leading
to the same values of the low-energy constants in the 
infinite volume give the same volume dependence.

Although it facilitates the comparison to chiral perturbation theory,
it is not necessary as a matter of principle to use the chiral
perturbation theory result for the mass dependence of the pion decay constant. 
However, as has been found for infinite volume, in order to correctly
describe the behavior of the pion decay 
constant as a function of a single symmetry breaking parameter, it is
necessary to go beyond the approximation of a constant expectation
value for the meson field, which we used in this paper, and to include wave
function renormalizations in the RG flow \cite{Jendges}. This
makes it possible to recover the correct prefactors of chiral logarithms 
in the framework of the renormalization
group. Such an approach is more powerful than the present
one, since in addition to the volume dependence, it predicts the
dependence of $m_\pi$ and $f_\pi$ on 
the symmetry breaking parameter $m_c$.
We stress that
even in such an approach, it remains necessary to fit the parameters
at the UV scale to the correct values of the low energy
constants. Thus, for example the value of the pion decay constant in
the chiral limit is not a prediction of the model, but a necessary
input to constrain its parameters.
The full set of RG-equations including the 
wave function renormalization and coupling constant renormalization
equations would reduce the input parameters to the four-fermion
coupling and the current quark mass at the UV-scale. In connection
with the symmetry breaking ansatz eq.~(\ref{eq:potansatz}), these
equations are more 
complicated and have not been worked out yet.

A limit on the possible values of the current quark mass is given by the 
requirement that all masses, in particular the sigma-mass, remain
substantially smaller than the  
ultraviolet cutoff $\Lambda_{UV} \approx 1500 \;\mathrm{MeV}$ of the
model. For a pion mass of $m_\pi=300\;\mathrm{MeV}$, we find
$m_\sigma\approx 800\;\mathrm{MeV}$. 

With regard to the UV-cutoff, we find
only a slight dependence of our results for reasonably large volumes. 
When we change the cutoff from
$\Lambda_{UV}=1500\;\mathrm{MeV}$ to
$\Lambda_{UV}=1100\;\mathrm{MeV}$, our results for the relative
shift $R[m_{\pi}(L)]$ of the pion mass in the finite volume 
change little.
The change in the pion mass from a variation of the cutoff is of the
order of less than $1\%$ for $L>2 \; \mathrm{fm}$, and approximately
$6\%$ at $L=1\; \mathrm{fm}$ for 
the largest pion mass we considered 
here, $m_\pi = 300 \;\mathrm{MeV}$. For smaller pion mass, the
dependence on the UV cutoff becomes weaker, for $m_\pi = 100
\;\mathrm{MeV}$ it is negligible on the scale of our results.
This can be understood, since a higher degree of
explicit symmetry 
breaking leads to 
more massive particles for which a smaller value for the UV momentum cutoff
becomes more relevant.

The sums over the momentum modes in the flow equations cannot be
performed analytically. 
For a numerical evaluation of the flow equations, these sums must be
truncated at a maximal mode number $N_{\mathrm{max}}=\mathrm{max}|\vec
n|$ which 
defines the cutoff momentum mode $p_{\mathrm{max}}=\frac{2 \pi}{L}
N_{\mathrm{max}}$. This numerical
truncation should not introduce an additional UV cutoff in
the model, and therefore we require that 
\be
\frac{2 \pi}{L} N_{\mathrm{max}} \gg \Lambda_{UV}.
\ee
Since we use a ``soft'' cutoff function it is necessary to really
satisfy the above equation with a safe margin.
For the volumes with $L \ge 1\;\mathrm{fm}$ considered here, 
we have used $N_{\mathrm{max}}=40$. 

%%%%%%%%%%%%%%%%%%%%%%%%%%%%%
\begin{figure}
\includegraphics[scale=0.80, clip=true, angle=0,
  draft=false]{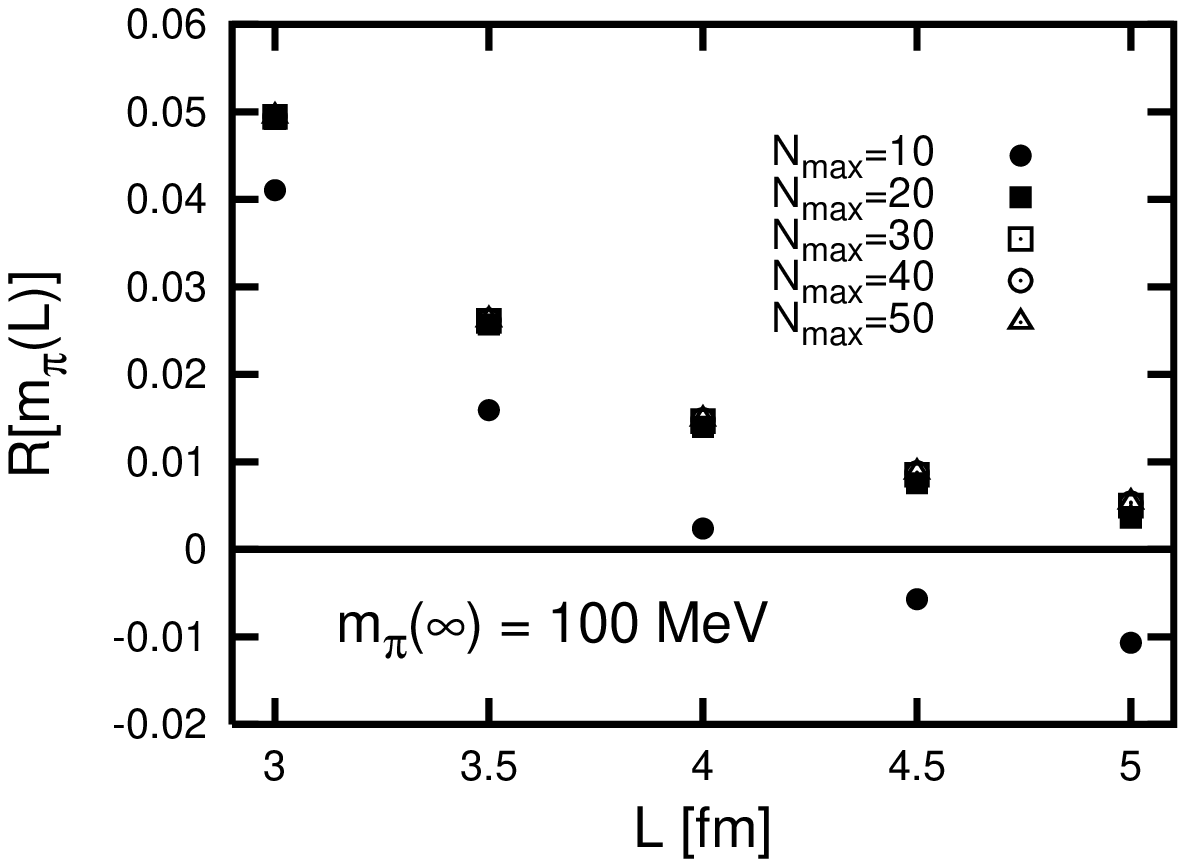}
\includegraphics[scale=0.80, clip=true, angle=0,
  draft=false]{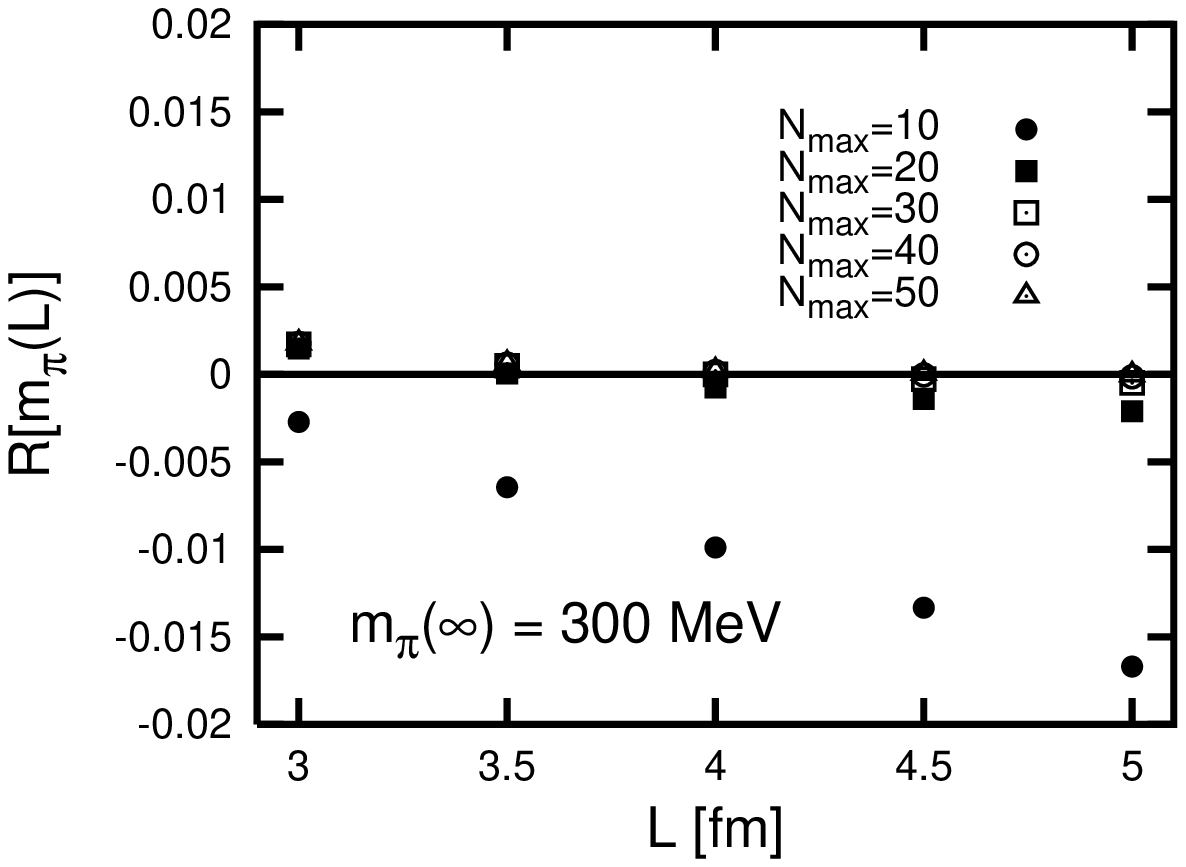} 
\caption{\label{fig:Nconvergence} The figures show the dependence of
  the results for the pion mass in finite volume on the truncation
  $N_{\mathrm{max}}$  of
  the momentum sum
  in the numerical evaluation of the flow equations. Plotted is
  $R[m_\pi(L)]$, the
  relative deviation of the finite volume pion mass from the value in infinite
  volume, as a function of the box size $L$, for different values of
  $N_{\mathrm{max}}$. Results for two different
  values of  
  the pion mass are shown in the two panels. Ideally, for large
  volumes, the relative deviation 
  from the infinite volume pion mass should approach zero. The
  truncation has a relatively larger effect for heavier pion
  masses.}
\end{figure}
%%%%%%%%%%%%%%%%%%%%%%%%%%

In figure~\ref{fig:Nconvergence}, the dependence of the relative
difference of the pion mass in finite volume from its value in the
infinite volume limit,  $R[m_\pi (L)]$, is shown as  
a function of $L$ for different values of the maximal mode number
$N_{\mathrm{max}}$. The relative mass difference depends mainly on 
$m_\pi L$ and drops exponentially for large  
values of this dimensionless variable. Thus, for any given value of $L$, the
value of $R[m_\pi(L)]$ will be smaller for a heavier pion. Comparing the
two panels in figure~\ref{fig:Nconvergence}, we see that although the
absolute values are smaller, the relative error due to the finite
number of momentum modes in the evaluation is larger for a
heavier pion. The reason is the increasing importance of the
non-zero momentum modes when the pion mass becomes larger at fixed box
size. Following the argument from \cite{Gasser:1987ah} outlined
above, if $1/m_\pi \gg L$, then the partition function is dominated by the
zero modes and effects from finite momentum modes present small
corrections. When the pion mass is increased,
the importance of the finite momentum modes grows and the
number of modes  has to be increased to
obtain results with the same level of accuracy. This argument can
be presented in a more formal way. The momentum sums
contributing to the flow equations are of the form
\be
\sum_{n_1, n_2, n_3} \frac{1}{k^2+ m_\pi^2 + \vec{n}^2\frac{4 \pi^2}{L^2}}
\ee  
where $\vec{n}^2=n_1^2+n_2^2+n_3^2$. For small values of the renormalization
scale $k$, the sum is dominated by the zero mode term 
\be
\frac{1}{m_\pi^2}+\frac{1}{m_\pi^2 + \frac{4 \pi^2}{L^2}} + \ldots.
\ee
If the box is sufficiently small, all terms with non-zero momentum are
suppressed by $1/L^2$, which 
acts as a large regulator. As
we increase the size of the box, so that $1/L \sim m_\pi$, 
the contributions of the nonzero momentum modes are of the
same size as the zero mode term. Therefore, effects from
a truncation of the momentum sums should be expected to appear already
at a smaller box size for large pion masses, which is exactly what we observe.

\section{Results}
\label{sec:results}

The results of the RG-flow equations for the evolution with the
infrared
cutoff scale $k$ give a picture of chiral symmetry breaking which 
reflects the formation of the quark condensate for higher momenta and
the effects of pion fluctuations at low scales.
%%%%%%%%%%%%%%%%%%%%%%%%%%
\begin{figure}
\includegraphics[scale=0.62, clip=true, angle=0,
  draft=false]{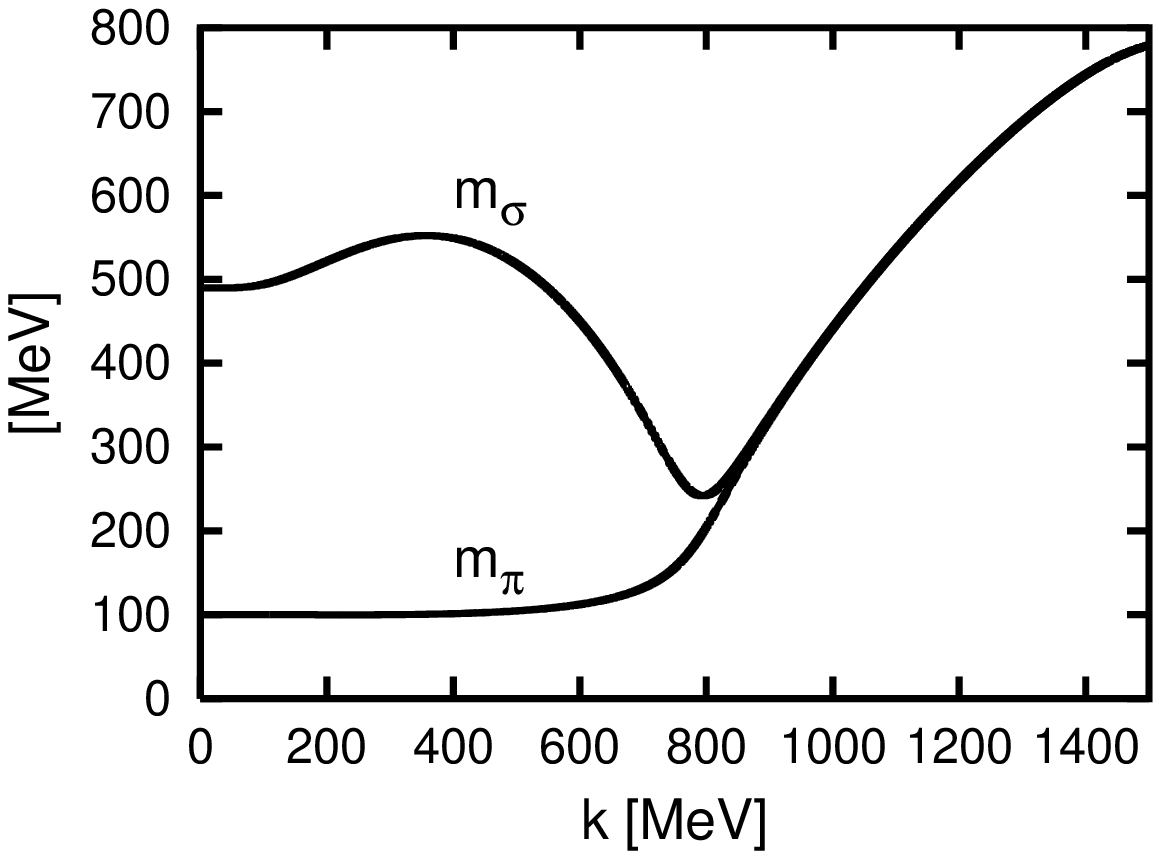} 
\includegraphics[scale=0.62, clip=true, angle=0,
  draft=false]{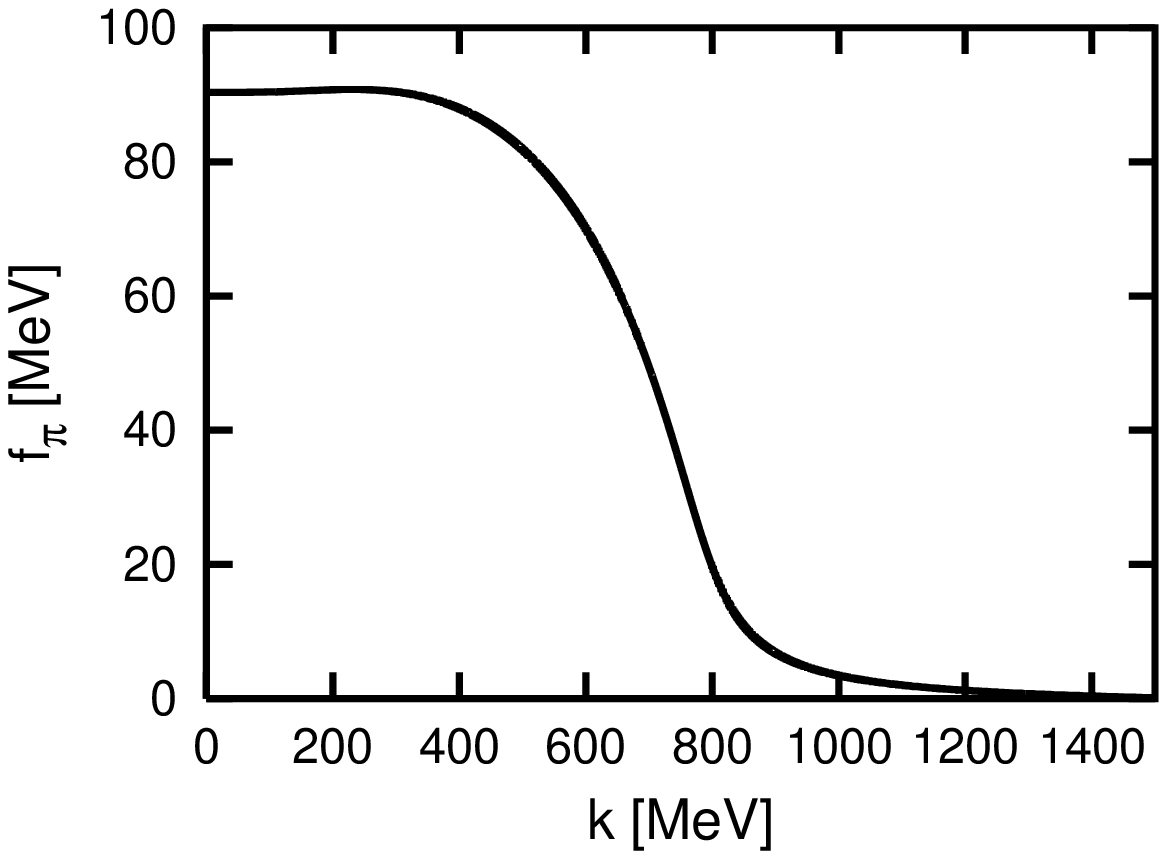}
\caption{\label{fig:flowk} Masses of the mesonic
  degrees of freedom and the pion decay constant as a function of the
  renormalization scale $k$ in infinite volume. The chiral symmetry
  breaking scale can be 
  clearly identified as the scale at which the mass of the heaviest meson
  (the $\sigma$) has a minimum. For this figure, we have chosen
  $m_\pi(\infty)=100 \;\mathrm{MeV}$ and $f_\pi(\infty)=90.4 \;\mathrm{MeV}$.}
%\end{figure}
%%%%%%%%%%%%%%%%%%%%%
%%%%%%%%%%%%%%%%%%%%%%%%%%
%\begin{figure}
\vspace{0.8cm}
\includegraphics[scale=0.62, clip=true, angle=0,
  draft=false]{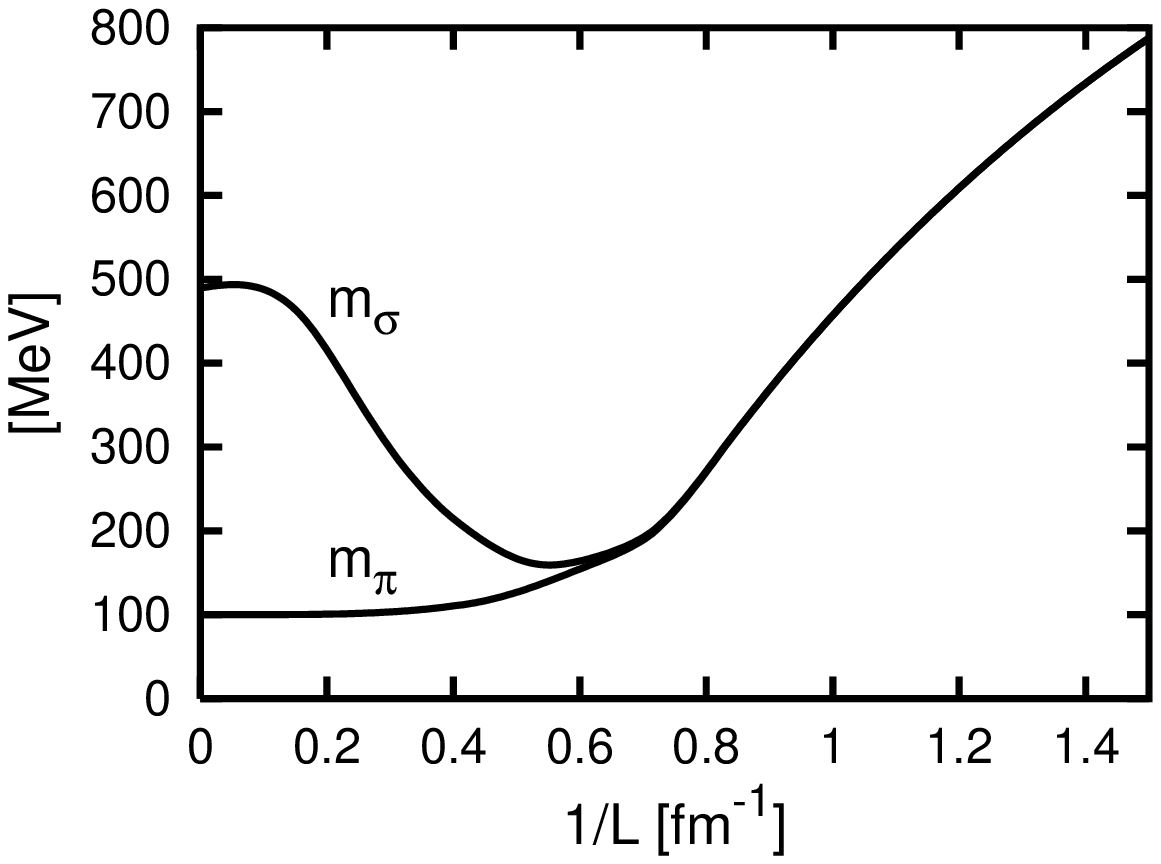} 
\includegraphics[scale=0.62, clip=true, angle=0,
  draft=false]{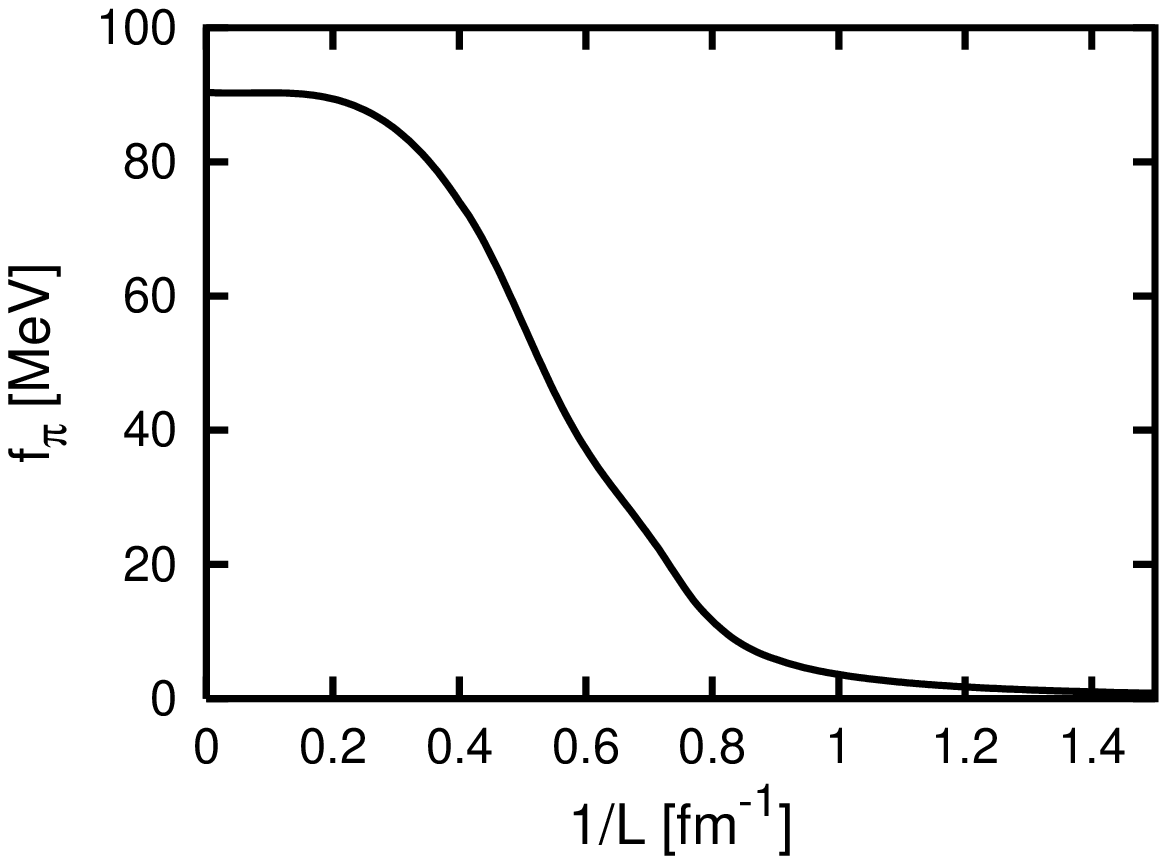}
\caption{\label{fig:flowL} Masses of the mesonic
  degrees of freedom and the pion decay constant as a function of the
  inverse box size $1/L$. The results are obtained by completely
  integrating out all quantum fluctuations ($k \to 0$) at fixed
  $L$. As soon as $k < 1/L$, the box size becomes the controlling
  scale, and in the limit 
  $k \to 0$ it is the only scale that remains. As for the
  preceding figures, we show the results with $m_\pi=100
  \;\mathrm{MeV}$ and $f_\pi=90.4 \;\mathrm{MeV}$ for $k \to 0$ and $L
  \to \infty$.} 
\end{figure}
%%%%%%%%%%%%%%%%%%%%%
Figure~\ref{fig:flowk} shows the masses of the pion and
sigma, and the 
pion decay constant as a function of the renormalization scale $k$,
for a value of $m_\pi = 100 \; \mathrm{MeV}$, in the infinite volume
limit. 
Starting at the UV scale $\Lambda_{UV}$ and proceeding towards smaller
values of $k$, we observe that the 
pion decay constant $f_\pi$ grows rapidly around the chiral symmetry
breaking scale 
$k_{\chi SB}\;\sim \; 800\;\mathrm{MeV}$, begins to flatten between
$600-400\; \mathrm{MeV}$, and becomes almost completely flat below
$300\;\mathrm{MeV}$. Generally, massive degrees of freedom decouple
from the renormalization group flow at a momentum scale given by the value
of their mass $m$, i.e. they do not contribute to the renormalization 
for $k < m$. This can be seen clearly 
in the flow of $f_\pi$. As soon as the
renormalization scale is of the order of the
constituent quark mass (approximately $300
\;\mathrm{MeV}$), the quarks are no longer dynamical degrees of
freedom and $f_\pi$ becomes essentially constant. 
The RG flow of the mass of the heaviest meson, the sigma, is in several
respects very similar to the flow of $f_\pi$. Its slope is also
initially large at the chiral 
symmetry breaking scale and starts to decrease between $600-400\;
\mathrm{MeV}$ as well. The value of the sigma-mass reaches a maximum
at $k$ slightly above 
$300\;\mathrm{MeV}$. Its decrease below this scale is due to the light
pion with a mass of 
$100\; \mathrm{MeV}$, which remains in the evolution 
as the only dynamical degree of freedom. 
When the pion
mass is increased, the drop in the sigma-mass below the scale set by the
constituent quark mass becomes much less pronounced. For
$m_\pi \approx m_q \approx 300\;\mathrm{MeV}$, $m_\sigma(k)$ 
is essentially a flat function of $k$
after it has reached its maximum.   

In finite volume, a similar behavior is visible.
In figure~\ref{fig:flowL}, the meson masses and the pion decay
constant are shown as a function of the scale $1/L$ set by the finite
volume. In these results, all quantum fluctuations
are integrated out completely, which removes the scale $k$.
Now let us consider a finite value of $k$, where the quantum
fluctuations are only partially integrated out. 
The scale $1/L$ introduced by the finite volume
is in competition with the renormalization scale
$k$. As soon as $k$ drops below $\pi/L$, the renormalization
scale no longer controls the renormalization
flow. We can interpret the results shown in
fig~\ref{fig:flowL} roughly as an instant picture of the k-flow arrested at a
scale $k=\pi/L$. 
However, this correspondence is not one-to-one: while
the cutoff $k$ affects both bosonic and fermionic fields in the same
way, this is not true for $1/L$. 
Since there are no zero modes for fields with anti-periodic
boundary conditions, the fermionic fields are more strongly affected
by this cutoff than the mesons. For the mesons, 
the scale $\frac{2 \pi}{L}$  imposes only 
a minimum value for the smallest non-zero momentum mode. For the
fermionic fields, on the other hand, the
lowest momentum mode $\sqrt{3} \pi/L$ can effectively
``freeze'' the quark fields already above the constituent quark mass
scale and no condensation of quarks takes place. 
For very small volumes $1/L> 0.5 \;\mathrm{fm}^{-1}$, the suppression of quark
condensation by the large cutoff becomes the dominating effect. 
The chiral symmetry is approximately restored. 
A more subtle effect can also be  seen in the
behavior of the sigma-mass. While the sigma-mass has a maximum
in the $k$-flow at a value of $m_\sigma \approx 600 \;\mathrm{MeV}$,
from which it drops to $m_\sigma \approx 500
\;\mathrm{MeV}$ due to the pion fluctuations, there is no
corresponding maximum in the $1/L$-dependence. With
decreasing $1/L$ the
pions hardly feel the constraints of the finite
volume, but the momentum scale at which the quarks decouple
consistently increases. Therefore the pion
contributions at low momenta have a
greater effect in the RG flow, since the $k$-region increases in
which they are the only relevant degrees of freedom. 
Through this effect, the pions also contribute toward the restoration
of chiral symmetry for small volumes.

The volume dependence of the low energy observables is
the main results of this paper. 
In table~\ref{tab:Rmpi}, we give the values for $R[m_\pi(L)]$,
cf. eq.~(\ref{eq:Luscher}), the relative difference
of the pion mass $m_\pi(L)$ in finite volume from 
$m_\pi(\infty)$, its value in infinite volume, for three volume sizes 
$L=2.0,\,2.5,\,3.0\,\mathrm{fm}$ and three pion masses $m_\pi(\infty)=100, 
  \, 200, \, 300\,\mathrm{MeV}$.    

\begin{figure*}[h!]
\hspace*{0.0cm}
\includegraphics[scale=0.80, clip=true, angle=0,
  draft=false]{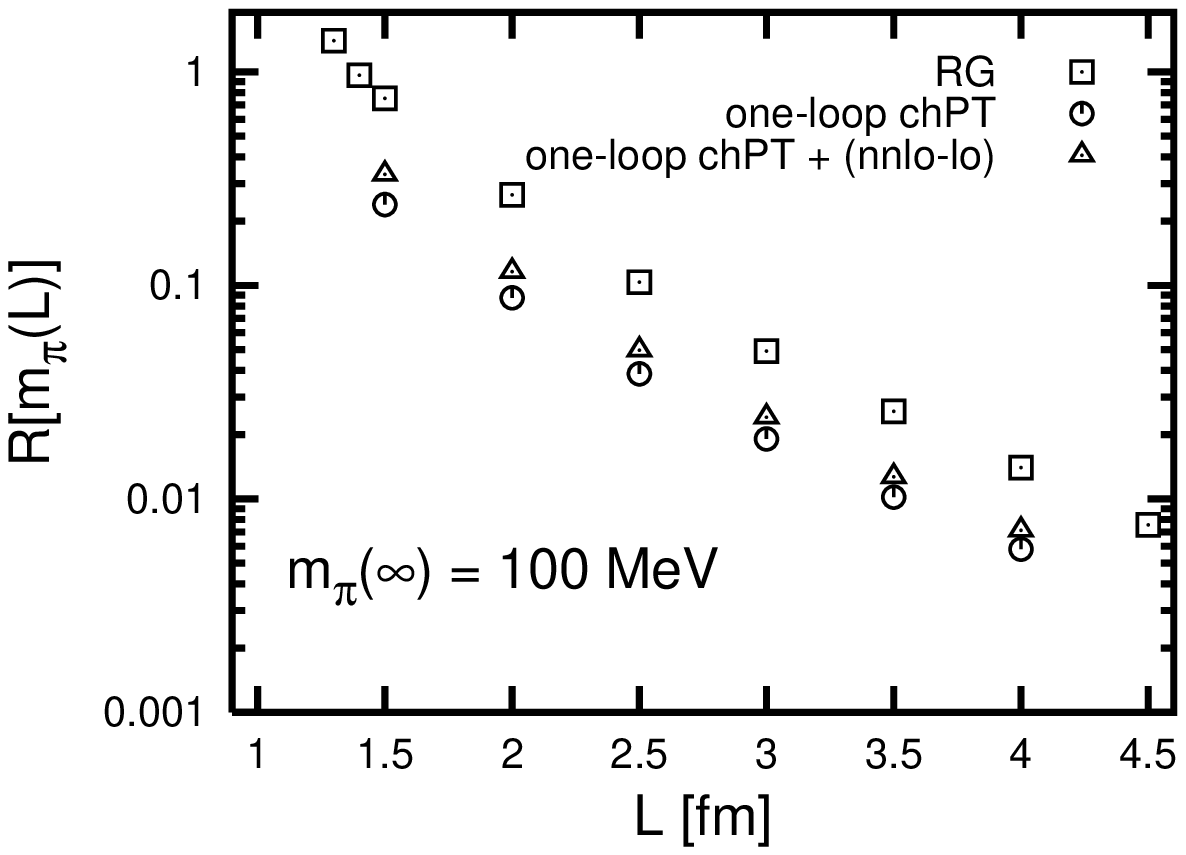} 
\includegraphics[scale=0.80, clip=true, angle=0,
  draft=false]{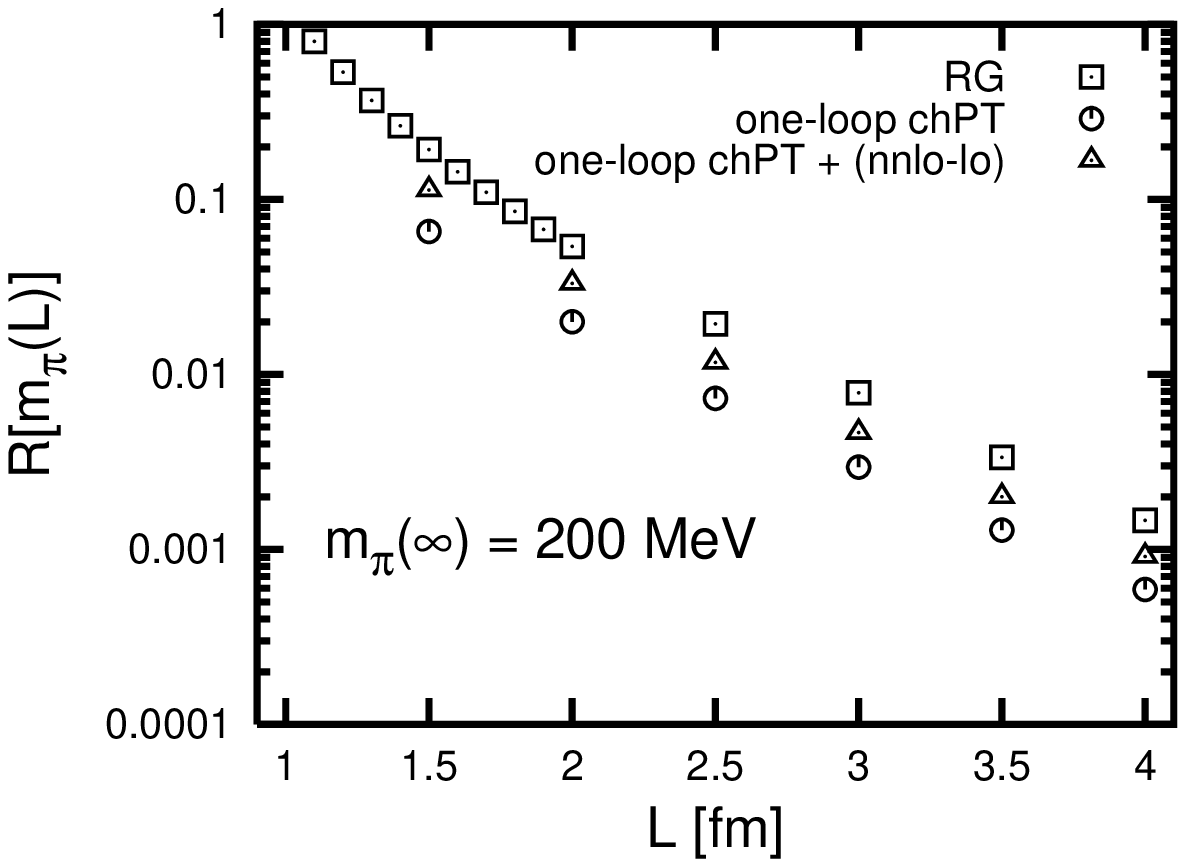} 
\includegraphics[scale=0.80, clip=true, angle=0,
  draft=false]{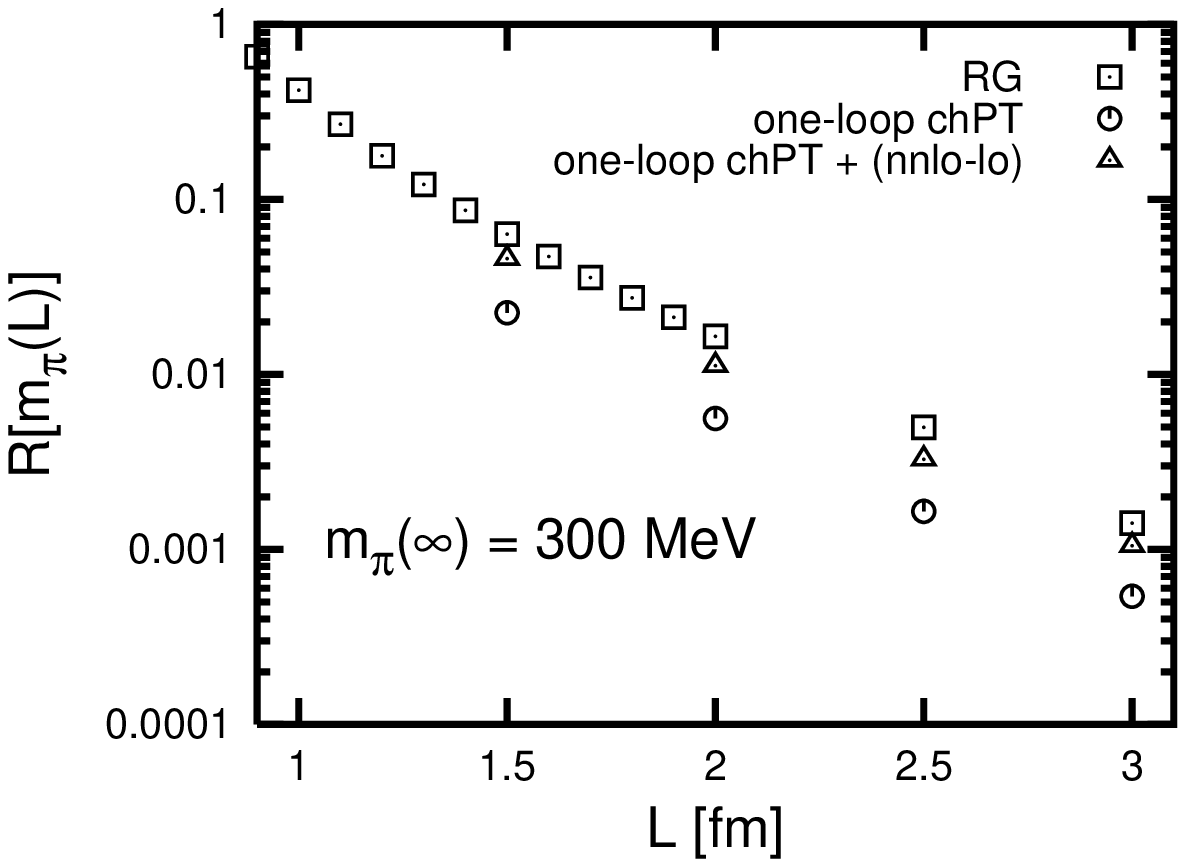} 
\caption{\label{fig:Rmpi} Volume dependence of the pion mass. We plot the
  relative shift of the pion mass from its infinite volume limit
  $R[m_\pi(L)] = (m_\pi(L)-m_\pi(\infty))/m_\pi(\infty)$ as a function
  of the size of the volume $L$. For comparison, we also
  plot the results from chPT calculations taken from
  \cite{Colangelo:2003hf}. The values for the pion
  mass in infinite volume are given in the figure, note the different
  scales on the axes for different $m_\pi(\infty)$.}
\end{figure*}

\begin{figure*}[h!]
\hspace*{0.0cm}
\includegraphics[scale=0.80, clip=true, angle=0,
  draft=false]{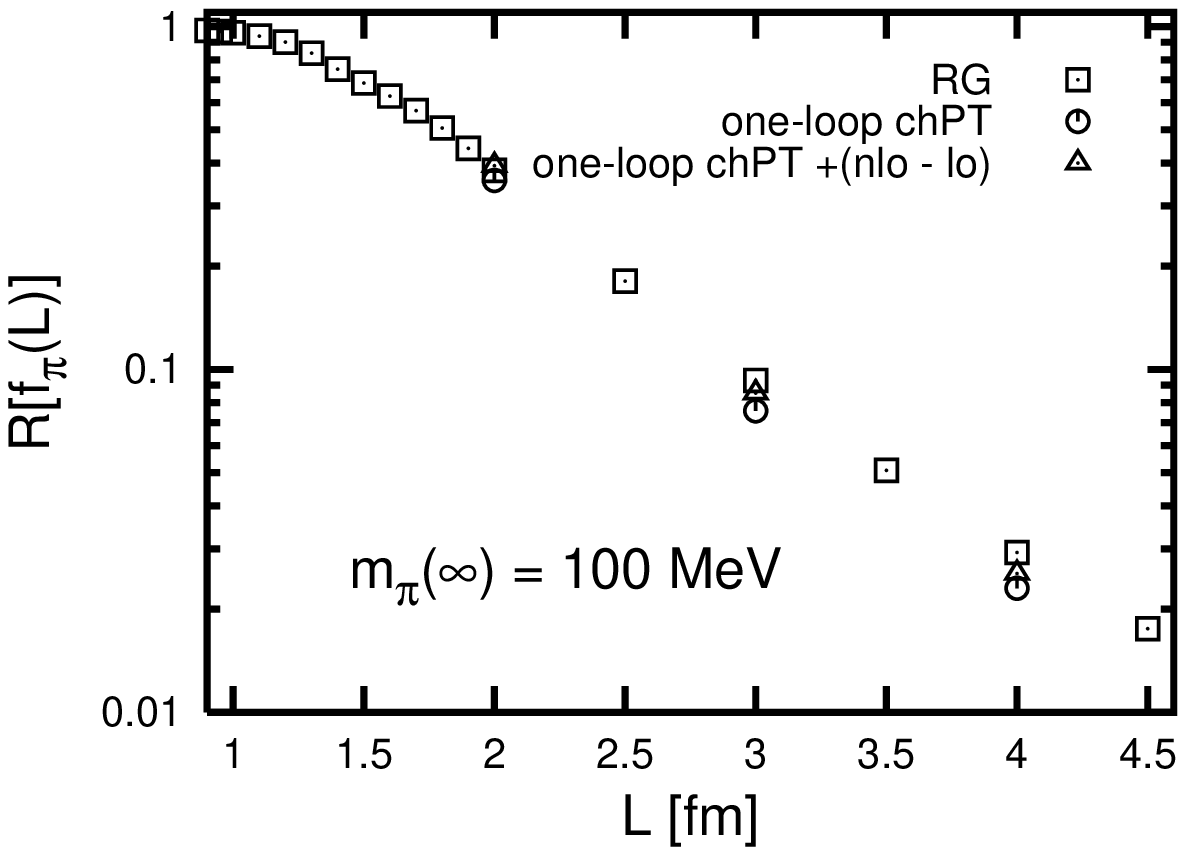} 
\includegraphics[scale=0.80, clip=true, angle=0,
  draft=false]{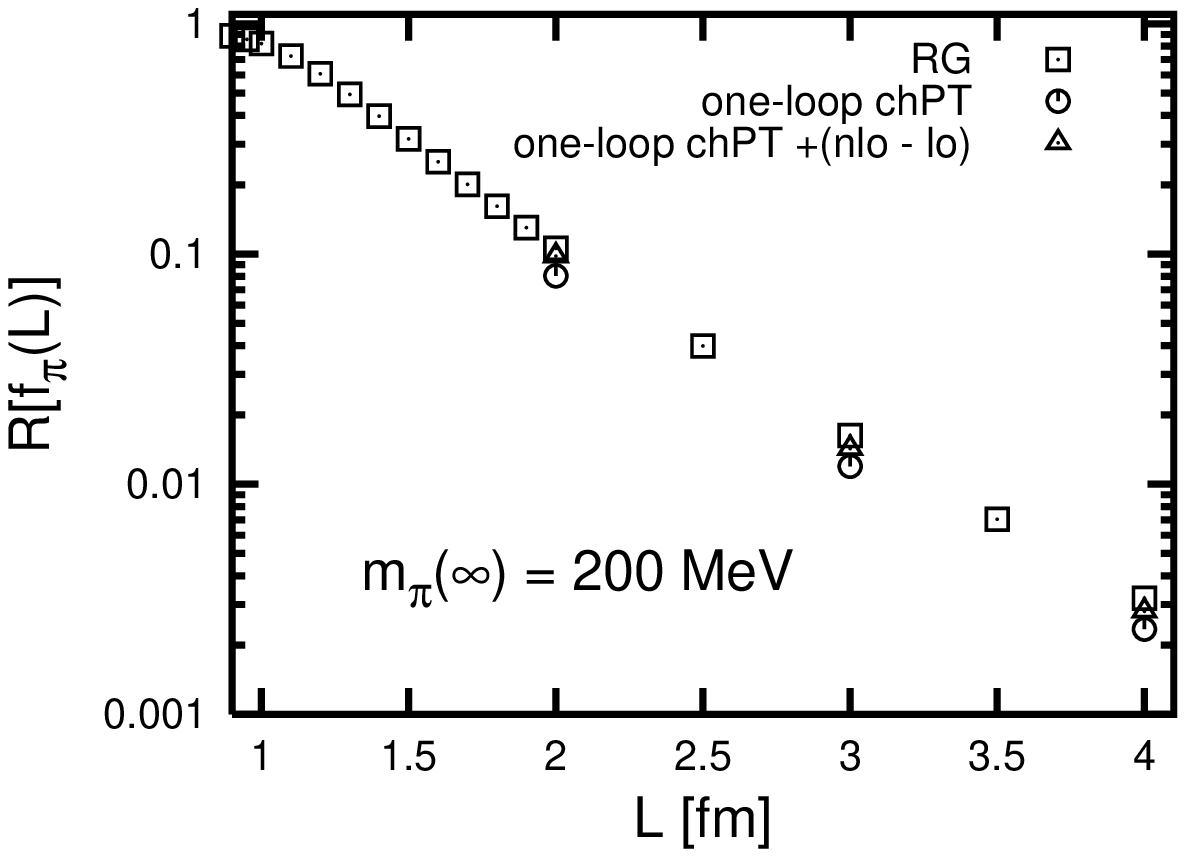} 
\includegraphics[scale=0.80, clip=true, angle=0,
  draft=false]{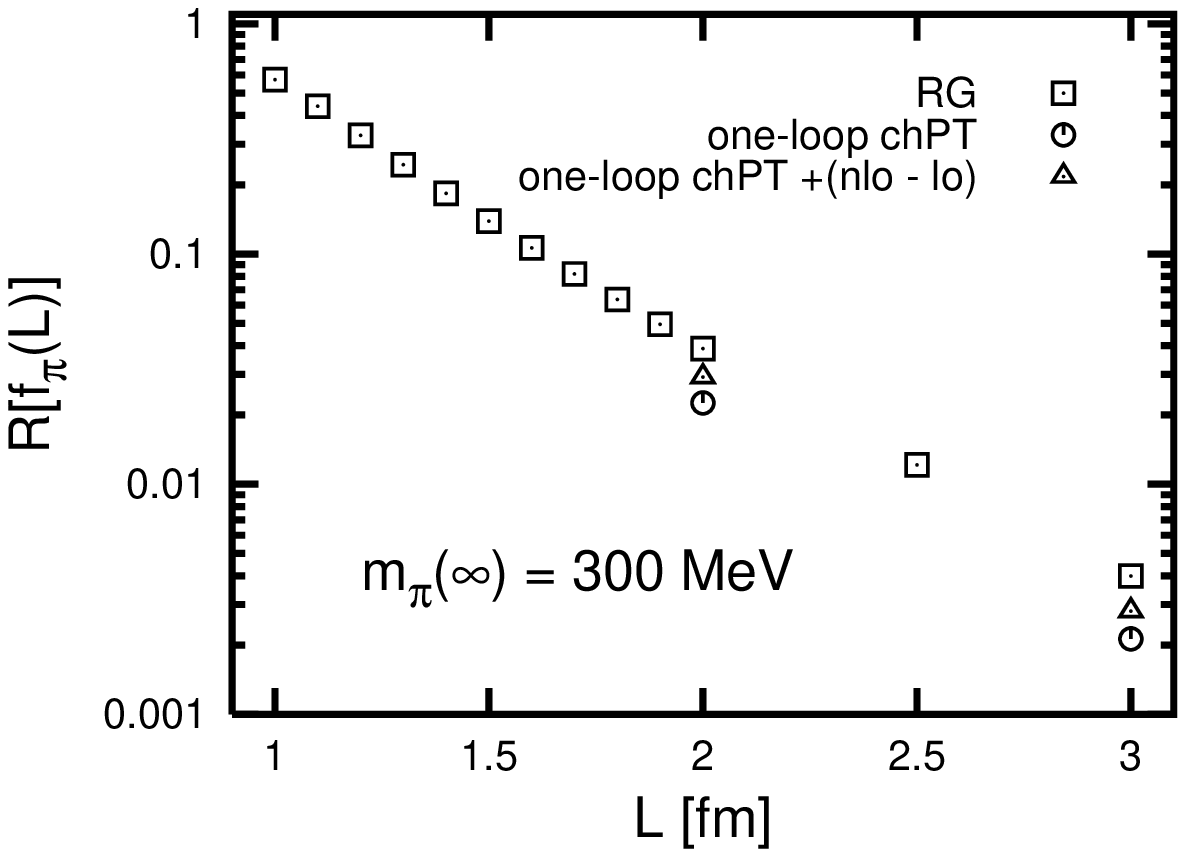} 
\caption{\label{fig:Rfpi} Volume dependence of the pion decay constant. We
  plot the relative shift from the infinite volume limit
  $R_{f_\pi}(L) = (f_\pi(\infty)-f_\pi(L))/f_\pi(\infty)$ as a
  function of the box size $L$ for
  different values of the pion mass. 
  For comparison, we show
  the chPT results taken from
  \cite{Colangelo:2004xr}. The values for $m_\pi(\infty)$ are given in the
  panels. Note the different scales on the axes for different $m_\pi(\infty)$.}
\end{figure*}

In figures~\ref{fig:Rmpi} and \ref{fig:Rfpi}, we show 
the relative change of the pion mass $m_\pi(L)$ and
the pion decay constant $f_\pi(L)$ as a function of the size $L$ of the
three-dimensional volume. We plot the results for the
relative differences on a logarithmic scale for
three different values of the pion mass 
$m_{\pi}(\infty)=100,\, 200,\, 300\,\mathrm{MeV}$.

%%%%%%%%%%%%%%%%%%%%%%%%%%%%%%%%%%%%%%%%%%%%%%%%%%%%%%%%%%%%%%%%%%%%%%%%%%
%%%%%%%%%%%%%%%%%%%%%%%%%%%%%%%%%%%%%%%%%%%%%%%%%%%%%%%%%%%%%%%%%%%%%%%%%%%
%%%%%%%%%%%%%%%%%%%%%%%%%%%%%%%%%%%%%%%%%%%%%%%%%%
\begin{table}
\begin{ruledtabular}
\begin{tabular}{cclrrrrr}
\multicolumn{3}{c}{ } &\multicolumn{3}{c}{${R[m_\pi(L)]}$}& \\
$L\;\mathrm{[fm]}$ & $m_\pi(\infty)\; \mathrm{[MeV]}$ & \hspace{0.31 em}
$m_\pi L$ & RG \hspace{1.8em} & 1L chPT \hspace*{0.3 em}& +({\it
  nnlo}-{\it lo}) \hspace{0.1em}& 
$\Delta R$\hspace{1.79em} \\ 
\hline
2.0 & 100 & 1.01293 & $26.6 \times 10^{-2}$ & $8.74 \times 10^{-2}$ &
$11.6 \times 10^{-2}$ &  $15.0 \times 10^{-2}$\\  
  & 200 & 2.02586 & $5.38 \times 10^{-2}$  & $2.00 \times 10^{-2}$ &
$3.31 \times 10^{-2}$ & $2.07 \times 10^{-2}$ \\ 
  & 300 & 3.03879 & $1.70 \times 10^{-2}$ & $0.56 \times 10^{-2}$&
$1.12 \times 10^{-2}$ & $0.58 \times 10^{-2}$ \\ 
\hline
2.5 & 100 & 1.26616 & $10.37 \times 10^{-2} $ & $3.85 \times 10^{-2}$
& $4.97 \times10^{-2}$ & $5.40 \times 10^{-2}$ \\ 
    & 200 & 2.53233 & $1.95 \times 10^{-2}$ & $0.73 \times 10^{-2}$ &
$1.17 \times 10^{-2}$ & $0.78 \times 10^{-2}$  \\ 
    & 300 & 3.79849 & $5.31 \times 10^{-3}$ & $1.65 \times 10^{-3}$ &
$3.27 \times 10^{-3}$ & $2.04 \times 10^{-3}$  \\  
\hline  
3.0 & 100 & 1.5194 & $4.94 \times 10^{-2}$ & $1.91\times 10^{-2}$ &
$2.41\times 10^{-2}$ & $2.53 \times 10^{-3}$  \\ 
    & 200 & 3.03879 & $7.85 \times 10^{-3}$ & $2.95 \times 10^{-3}$ &
$4.65 \times 10^{-3}$ & $3.20 \times 10^{-3}$ \\  
    & 300 & 4.55819 & $1.76 \times 10^{-3}$ & $0.54 \times 10^{-3}$ &
$1.05 \times 10^{-3}$ & $0.71 \times 10^{-3}$  \\ 
\end{tabular}
\end{ruledtabular}
\caption{\label{tab:Rmpi}Values for $R[m_\pi(L)]$,
  cf. eq.~(\ref{eq:Luscher}), the relative shift 
  of the pion mass in finite volumes of
  $L=2.0,\,2.5,\,3.0\,\mathrm{fm}$, compared to the value in
  infinite volume, for 
  pion masses of $m_\pi(\infty)=100, 
  \, 200, \, 300\,\mathrm{MeV}$.  We compare our RG
  calculation to the exact one-loop chPT results of
  \cite{Gasser:1987ah} for a finite volume (1L chPT), and
  the exact one-loop calculation with corrections in
  three-loop order obtained with chPT using L\"uscher's formula
  \cite{Colangelo:2003hf} (1L chPT + ({\it nnlo}-{\it lo})). In the last
  column, the difference $\Delta R$ between the RG result and the
  three-loop corrected chPT result is given.} 
\end{table}
%%%%%%%%%%%%%%%%%%%%%%%%%%%%%%%%%%%%%%%%%%%%%%%%%%%%%%%%%%%%%%%%%%%%%%%%%%%%%% 

Let us first discuss the plots for the pion mass in fig.~\ref{fig:Rmpi}. 
The relative change of the pion mass decreases with the volume size
$L$ and the pion mass $m_\pi(\infty)$. 
Fig.~\ref{fig:Rmpi} also contains the results of chiral perturbation theory 
from the exact one-loop calculation in finite
volume
\cite{Gasser:1986vb} and also the ``best estimate'' from
\cite{Colangelo:2003hf}. As discussed in section~\ref{sec:chPT}, the 
difference between the results of L\"uscher's formula for the mass
shift which uses as input $\pi \pi$-scattering 
in the one-({\it lo}) and three-({\it nnlo}) loop order is 
used as a correction 
to the {\it exact} one-loop mass shift from \cite{Gasser:1986vb}.

Our RG results have the same slope as those from chiral perturbation theory,
but are consistently above chiral
perturbation theory, even with corrections to three loops.
In general, the RG calculation gives values for $R[m_\pi(L)]$ which
are about a factor $1.5$ to $2.0$ larger than the values from chPT,
as can be seen in table~\ref{tab:Rmpi}.

The difference between the RG result and the loop expansion
decreases with higher order in loops. The RG result
is closer to the calculation with {\it nnlo}-L\"uscher formula than
to the one-loop chPT calculation.

For large volumes, the pion mass should drop as $\exp(-m_\pi
L)$. 
Therefore, we expect that the slope of the RG result is the same as
that from chiral perturbation theory, which is indeed the case.
The calculation with the flow equation can be continued to 
smaller box sizes, as long as the momenta constrained by the box are below the
cutoff $\Lambda_{UV}$. 
The slope of the RG result at small box size ($L < 1\, \mathrm{fm}$)
is approximately given by the meson mass $m_\sigma = m_\pi$,
cf. Fig.~\ref{fig:flowL}, after the transition between the 
regime dominated by chiral symmetry breaking and the one with restored chiral
symmetry has taken place.
In this other region the chiral expansion  
can no longer be considered reliable and is therefore no longer applied.

While the relative difference between the exact
one-loop result and the RG results remains approximately constant for
different pion masses, the relative difference between the RG 
results and the results of  Colangelo and D\"urr 
\cite{Colangelo:2003hf} decreases
when the mass of the pion is increased.
We have checked that the difference between our RG results and the
chiral perturbation theory results from Colangelo and D\"urr
are consistent with the error estimate of
L\"uscher's approximation formula. We find for all pion masses that
the differences decrease exponentially according to $\exp(- C\; m_\pi
L)$, with $C$ a positive constant of order $1$.

%%%%%%%%%%%%%%%%%%%%%%%%%%%%%%%%%%%%%%%%%%%%%%%%%%%%%%%%%%%%%%%%%%%%%%%%%%%
%%%%%%%%%%%%%%%%%%%%%%%%%%%%%%%%%%%%%%%%%%%%%%%%%%%%%%%%%%%%%%%%%%%%%%%%%%%

Next we discuss the results for the volume 
dependence of the pion decay constant
shown
in fig.~\ref{fig:Rfpi}. As for the pion mass, we compare with the chiral
perturbation theory results \cite{Gasser:1986vb} and \cite{Colangelo:2004xr}.
Note that in this case, we define the relative difference with
opposite sign as $R[f_\pi(L)] =
(f_\pi(\infty) - f_\pi(L))/f_\pi(\infty)$, because
the pion decay constant is {\it smaller} for a finite 
volume, in contrast to the pion mass. 
We refer to fig.~\ref{fig:flowL} for an illustration of the global
behavior of $f_\pi(L)$. When the size of the volume is decreased, the
pion decay constant at first drops only slowly, and then sharply at the
scale associated with chiral symmetry breaking.
For very small volumes, the pion decay
constant vanishes and chiral 
symmetry is effectively restored. This is true
for $L \rightarrow 0$ regardless of the value of the pion mass.
Therefore, the largest possible relative shift for $L \to 0$ is
$R[f_\pi(0)] =1 $, when the order parameter vanishes completely.

The plots for the volume dependence of $f_\pi$ 
illustrate in which region chiral perturbation theory
remains valid. Since spontaneously broken 
chiral symmetry and a sufficiently large $f_\pi$ are required, chPT
does not describe the transition to the region with effectively
restored chiral symmetry. Therefore, chiral perturbation theory
results are available only for volumes with
$L \ge 2\;\mathrm{fm}$. Already for $L= 2 \;\mathrm{fm}$ and a pion
mass $m_\pi=100 \;\mathrm{MeV}$, the $f_\pi$-shift is almost $40 \%$.
The RG method remains valid in both regions, down to very small volumes.

We compare with the chiral perturbation theory results obtained by using
an approximation similar to L\"uscher's formula for the pion mass shift,
which was derived in \cite{Colangelo:2004xr}. For
the pion decay constant, the input needed to calculate the finite volume shift
is an amplitude involving the axial current in infinite volume. 
So far only two loops in chiral perturbation theory ({\it nlo})
are known. The chPT {\it nlo}-corrections
to the exact one-loop result increase the shift towards larger values.
As in the case of the pion mass, the RG results give a slightly
larger finite volume shift than the chPT results. 
Again the results from RG and chPT converge for larger values of
the pion mass (note the different scales on both axes in
the plots for different values of $m_\pi$!).

For the volume dependence of the sigma-mass, we refer back to
fig.~\ref{fig:flowL}, where we show the overall dependence of
$m_\sigma$ on $1/L$. It is interesting to relate the variation of
$m_\sigma(1/L)$ to the $0^{++}$ phase shift of $\pi \pi$-scattering,
which will be done in a separate work.

\section{Conclusions}
\label{sec:conclusions}

We have presented a new approach to the quark-meson-model employing the
renormalization group method in a finite volume within the framework of
the Schwinger proper time formalism. 
Central to any such
approach is the inclusion of explicit chiral symmetry breaking. Since
chiral symmetry is not broken spontaneously in a finite volume, it is
necessary to introduce a finite current quark mass. 
In this paper, we have evolved the effective potential 
with
additional symmetry-breaking terms. The form of these terms is
constrained by the quark contributions to the renormalization group
flow, which introduce the explicit chiral symmetry breaking.

By solving the resulting renormalization group flow equations
numerically, we have obtained results for the volume dependence of the
meson masses, in particular the pion mass, and the pion decay
constant, the order parameter of chiral symmetry breaking.

Our results show consistently a larger finite volume mass shift for
the pion than
has been obtained in chiral perturbation theory
including up to three loops. 
The differences between the chiral perturbation theory results which
make use of the L\"uscher formula and the RG results obtained
in the present paper 
are consistent with the error estimate for L\"uscher's approximation. As
one expects, the difference is largest for small values of $m_\pi L$. We
have checked that this difference decreases exponentially with an
increase in this 
dimensionless quantity. As shown in figs.~\ref{fig:Rmpi},
\ref{fig:Rfpi}, our 
results and those obtained in chiral perturbation theory 
with  L\"uscher's formula \cite{Colangelo:2004xr,
  Colangelo:2003hf} converge for large
current quark masses.  
We note that the ratio of the results from chPT and RG does not depend
on $L$, even down to $L=1.5 \;\mathrm{fm}$. 

Compared to the present numerical approach, chiral perturbation theory
has the advantage that it is possible to
obtain analytical expressions for the finite volume mass shift. This makes
comparisons to lattice results simpler. On the other hand,
current lattice volumes and lattice pion masses are at
the edge of a reliable chiral perturbation theory calculation.
In contrast, the RG method remains valid for large current quark masses as
well as small volumes.
 
The main uncertainty of the RG method comes from its dependence 
on the UV cutoff scale $\Lambda_{UV}$ for large meson masses.
The system becomes sensitive to $\Lambda_{UV}$ for large explicit symmetry
breaking, because the mass of the sigma as the
heaviest particle approaches the UV cutoff.
For a pion mass of $m_\pi=300 \;\mathrm{MeV}$, the sigma mass is
$m_\sigma \approx 800\, \mathrm{MeV}$. In this case, a cutoff variation
between $\Lambda_{UV}=1500\;\mathrm{MeV}$ and $1100\;\mathrm{MeV}$,
changes the pion mass for a volume with $L=1\;\mathrm{fm}$ by
approximately $6\%$, and by less than $1\%$ for $L > 2
\;\mathrm{fm}$.
Within this uncertainty, our results agree with those of chPT for
$m_\pi = 300 \;\mathrm{MeV}$. 
In contrast, for $m_\pi=100 \;\mathrm{MeV}$ the cutoff dependence is so
weak that it is not noticeable on the scale of the results. The RG
and chPT results do not agree within this uncertainty,
cf. Table~\ref{tab:Rmpi} for $m_\pi=100\,\mathrm{MeV}$ and
$m_\pi=200\,\mathrm{MeV}$.  

The dependence of our results on the choice of model parameters at the
UV scale is much weaker than that on the cutoff. 
By fitting to the values of the low-energy observables  $m_\pi$ and
$f_\pi$ in infinite volume, we achieve a very high degree of
independence on the particular choice of UV parameters.

We expect that the inclusion of wave function renormalizations in the
finite volume renormalization group flow should make it possible
to describe the low energy constants as a function of a single
symmetry breaking parameter, the quark mass $m_c$. It has already been
observed that this is the case in infinite volume,
where the behavior of $m_\pi$ and $f_\pi$ as a function of $m_c$ is
described correctly when all other $UV$ parameters remain fixed
\cite{Jendges}. Systematic errors introduced by the simplification
used in this paper to adjust both $m_\pi(\infty)$ and $f_\pi(\infty)$ in
infinite volume could then be estimated by a comparison of the calculations.

Although we have not yet investigated these questions in depth, it appears
that the present approach, which treats the pion fields and the sigma
explicitly and as individual degrees of freedom, improves the
convergence \cite{Papp:1999he} of the polynomial expansion of the effective
potential. This is due to the finite mass acquired by the Goldstone
bosons, which becomes relevant for $k \to 0$.

The RG approach shows in a transparent way the relevance of the
momentum scale introduced by the finite volume for the quantum fluctuations.

\acknowledgments
H.J.P. would like to thank Prof. Leutwyler and Dr. D\"urr for instructive
discussions. J.B. would like to thank the GSI for financial support.

\appendix

\section{Flow equations}

We derive flow equations for the coefficients of the potential
by inserting eq.~(\ref{eq:potansatz}) into the flow equation for the potential and
comparing coefficients of both sides. 
We define expansion coefficients for the flow equations: 
\be
\left(k \frac{\partial}{\partial k} U_k \right)^{ij} &:=&\left. \frac{1}{i!}
\frac{1}{j!} \left(\frac{ \partial}{\partial \sigma}\right)^i
\left(\frac{\partial}{\partial \vec{\pi}^2} \right)^j
k \frac{\partial}{\partial k} U_k
\right|_{{ \sigma=\sigma_0 \atop \vec{\pi}^2 = 0}}. 
\ee
Then the flow equations for the coefficients $a_{ij}$ in the ansatz
for the potential are given by
\be
\left(k \frac{\partial}{\partial k} U_k \right)^{ij} &=& \left( k
\frac{\partial }{\partial k} a_{ij}\right) + a_{i+1, j} \left(- k
\frac{\partial \sigma_0}{\partial k} \right) (i+1) (1 -
\delta_{N_\sigma, i}) \el
&&+ a_{i, j+1} (j+1) \left(- 2 \sigma_0 k
\frac{\partial \sigma_0}{\partial k} \right) (1 -
\delta_{\frac{1}{2}(N_\sigma-i), j} )
\label{eq:flowcoeff}
\ee
with the additional condition that $(1+ 2j) \le N_\sigma$.

In order for $\sigma=\sigma_0$ to actually correspond to a
minimum of the potential, it must satisfy $\frac{\partial}{\partial \sigma}
\left.U_k \right|_{\sigma = \sigma_0} = 0$. 
For the coefficients $a_{10}$ and $a_{01}$, this translates into the
condition 
\be
a_{10} + 2 a_{01} \sigma_0 \equiv 0.
\label{eq:mincon}
\ee
Due to this condition, only two of the variables in the set
$\{a_{10}, a_{01}, \sigma_0\}$ are independent, and the third one can
be expressed in terms of the other two. Likewise, if we take the
derivative of the equation (\ref{eq:mincon}) with respect to the
renormalization scale $k$, we get an equation which relates the flow 
of these three variables:
\be
k \frac{\partial}{\partial k} a_{10} + 2 a_{01} k
\left (\frac{\partial}{\partial k} \sigma_0 \right )+ 2 \sigma_0 k
\frac{\partial}{\partial k} a_{01} = 0.
\label{eq:minflow}
\ee
We can use this equation to replace the flow
equation for $a_{10}$ in the above set
eq.~(\ref{eq:flowcoeff}). It is desirable to eliminate $a_{10}$, since
$\sigma_0$ and $a_{01}$ both 
correspond to the observables we wish to obtain, namely the pion decay
constant and the pion mass. In addition, with this replacement
the system of differential equations can also be solved more easily.

From the general expression for the flow equations
eq.~(\ref{eq:flowcoeff}), we find the particular equations governing
$a_{10}$ and $a_{01}$:
\be
\left(k \frac{\partial}{\partial k} U_k \right)^{10} &=& k
\frac{\partial}{\partial k} a_{10} - 2 a_{20}  \left( k
\frac{\partial}{\partial k} \sigma_0 \right ) - a_{11} \left( 2
\sigma_0 k \frac{\partial}{\partial k} 
\sigma_0 \right) \\
\left(k \frac{\partial}{\partial k} U_k \right)^{01} &=& k
\frac{\partial}{\partial k} a_{01} - a_{11} \left (k
\frac{\partial}{\partial k} 
\sigma_0 \right ) - 2 a_{02} \left( 2 \sigma_0 k
\frac{\partial}{\partial k} \sigma_0\right ).
\ee
The flow equation for $a_{10}$ contains on the LHS only terms that are
proportional to the symmetry-breaking current quark mass
$m_c$. Because of this, $a_{10}$ does not evolve in the
chiral limit $m_c \to 0$. If it is initially zero at the UV scale, it
remains zero on all scales.
In this case, the condition~(\ref{eq:minflow}) forces the coefficient
$a_{01}$ to vanish as soon as $\sigma_0$ acquires a finite expectation
value. This corresponds to the appearance of exactly massless
Goldstone bosons in case of spontaneous symmetry breaking, in
accordance with our expectations for the chiral limit. 

In order to derive a flow equation for the minimum of the potential
$\sigma_0$, we can combine the two equations and use
eq.~(\ref{eq:minflow}) to eliminate the $k$-derivatives of $a_{10}$
and $a_{01}$:
\be
\left(k \frac{\partial}{\partial k} U_k \right)^{10} + 2 \sigma_0
\left(k \frac{\partial}{\partial k} U_k \right)^{01} &=& -
\left (k \frac{\partial}{\partial k} \sigma_0 \right ) \left(2 a_{20}
+ 2 a_{01} + 4 a_{11} \sigma_0 + 8 a_{02} \sigma_0^2\right).
\ee
From the expressions for the meson masses, evaluated at the minimum of
the potential, it can be seen that the expression in
brackets, which multiplies the $k$-derivative of $\sigma_0$, is up to
a constant factor the square of
the $\sigma$-mass, $M_\sigma^2$. Therefore, this equation is always
well-conditioned. The only exception is at the chiral symmetry
breaking scale, where $M_\sigma^2$ drops sharply, if the explicit symmetry
breaking is very small. For reasonably large pion masses, this is
not a problem.

\section{Ansatz for the potential}

We use the flow equations in infinite volume to motivate the
ansatz for the  potential with explicit symmetry breaking.  
Neglecting the mesonic contributions to the flow equations, we are
left with the terms arising from the fermions:
\be
k \frac{\partial}{\partial k} U^{F}(\sigma, \vec{\pi}^2) &=& -
\frac{ 4 N_f N_c}{32 \pi^2} \frac{k^6}{k^2 + M_q^2}.
\ee
The constituent quark mass $M_q$ contains through
the current quark mass $m_c$ the only explicitly symmetry breaking
term in the flow equation. As shown in eq.~(\ref{eq:mqexpansion}), by
expanding around the minimum of the 
potential, the quark mass can be written as
\be
M_q^2 &=&g^2 [(\sigma_0 + m_c)^2 + 2 m_c (\sigma - \sigma_0) +
(\sigma^2 +\vec{\pi}^2 -\sigma_0^2)],
\ee
where $m_c$ is rescaled by a factor of $g$ for
convenience. When we expand the denominator in the flow equation in the
deviation of the fields from the vacuum expectation value, the result
contains only those terms we postulated in our ansatz 
for the potential:
\be
k \frac{\partial}{\partial k} U^{F}(\sigma, \vec{\pi}^2) &=& -
\frac{1}{32 \pi^2} 4 
N_f N_c \frac{k^6}{k^2 + g^2 (\sigma_0+m_c)^2}\times \el
&& \times  \Bigg\{ (\sigma^2+\vec{\pi}^2 -\sigma_0^2)\;\;\left[-
  \frac{g^2}{k^2+g^2(\sigma_0+m_c)^2} \right] \el 
&&\quad  +(\sigma^2+\vec{\pi}^2 -\sigma_0^2)^2
\left[\left(\frac{g^2}{k^2+g^2(\sigma_0+m_c)^2}\right)^2 \right] \el 
&&\quad +(\sigma-\sigma_0)\; \left[- 2
  m_c\left(\frac{g^2}{k^2+g^2(\sigma_0+m_c)^2}\right) \right] \el 
&&\quad +(\sigma-\sigma_0)^2 \left[4 m_c^2
  \left(\frac{g^2}{k^2+g^2(\sigma_0+m_c)^2}\right)^2 \right]\el 
&&\quad +(\sigma-\sigma_0) (\sigma^2+\vec{\pi}^2 -\sigma_0^2)\; 2\;
\left[- 2 m_c \left(\frac{g^2}{k^2+g^2(\sigma_0+m_c)^2}\right)^2 \right]\el
&&\quad + \ldots
\Bigg\}.
\ee
All remaining terms are of higher order in $(\sigma-\sigma_0)$
or $(\sigma^2+\vec{\pi}^2 - \sigma_0^2)$ or any combination thereof.
Only the terms in
$(\sigma^2+\vec{\pi}^2 -\sigma_0^2)=(\phi^2-\sigma_0^2)$ remain in the 
chiral limit $m_c \to 0$. These terms respect the chiral
symmetry and the potential reduces to the symmetric form.

\end{document}